\documentclass[prl,aps,twocolumn,superscriptaddress,nofootinbib,preprintnumbers,floatfix]{revtex4-1}
\pdfoutput=1

\usepackage{amsmath}
\usepackage{graphicx}
\usepackage[small]{subfigure}
\usepackage[hyperfootnotes=false]{hyperref}
\usepackage{xspace}

\newcommand{\eq}[1]{Eq.~\eqref{eq:#1}}
\newcommand{\eqs}[2]{Eqs.~\eqref{eq:#1} and \eqref{eq:#2}}
\newcommand{\fig}[1]{Fig.~\ref{fig:#1}}

\newcommand{\ord}[1]{\mathcal{O}(#1)}

\newcommand{\df}{\mathrm{d}}
\newcommand{\img}{\mathrm{i}}

\newcommand{\tr}{\mathrm{tr}}
\newcommand{\sdt}{\!\cdot\!}

\newcommand{\al}{\alpha}
\newcommand{\de}{\delta}

\newcommand{\cB}{{\mathcal B}}
\newcommand{\cE}{{\mathcal E}}
\newcommand{\cI}{{\mathcal I}}

\newcommand{\cY}{{\mathcal Y}}

\newcommand{\nn}{\nonumber}

\newcommand{\Ecm}{E_\mathrm{cm}}
\newcommand{\lqcd}{\Lambda_\mathrm{QCD}}

\newcommand{\GeV}{\mathrm{GeV}}
\newcommand{\TeV}{\mathrm{TeV}}

\newcommand{\Pythiaeight}{\textsc{Pythia}8\xspace}
\newcommand{\Herwig}{\textsc{Herwig++}\xspace}

\allowdisplaybreaks[4]

\begin{document}

%%%%%%%%%%%%%%%%%%%%%%%%%%%%%%%%%%%%%%%%%%%%%%%%%%%%%%%%%%%%%%%%%%%%%%%%%%%%%%%%
% Title page
%%%%%%%%%%%%%%%%%%%%%%%%%%%%%%%%%%%%%%%%%%%%%%%%%%%%%%%%%%%%%%%%%%%%%%%%%%%%%%%%

\preprint{\vbox{
\hbox{MIT--CTP 4530}
\hbox{DESY 14-008}
\hbox{NIKHEF 2014-002}
}}

\title{Dissecting Soft Radiation with Factorization}

\author{Iain W.~Stewart}
\affiliation{Center for Theoretical Physics, Massachusetts Institute of Technology, Cambridge, MA 02139, USA\vspace{0.5ex}}

\author{Frank J.~Tackmann}
\affiliation{Theory Group, Deutsches Elektronen-Synchrotron (DESY), D-22607 Hamburg, Germany\vspace{0.5ex}}

\author{Wouter J.~Waalewijn\vspace{1.ex}}
\affiliation{Nikhef, Theory Group, Science Park 105, 1098 XG, Amsterdam, The Netherlands
\vspace{0.5ex}}
\affiliation{ITFA, University of Amsterdam, Science Park 904, 1018 XE, Amsterdam, The Netherlands
\vspace{0.5ex}}

\date{May 26, 2014}

%%%%%%%%%%%%%%%%%%%%%%%%%%%%%%%%%%%%%%%%%%%%%%%%%%%%%%%%%%%%%%%%%%%%%%%%%%%%%%%%
\begin{abstract}

An essential part of high-energy hadronic collisions is the soft hadronic activity that underlies the primary hard interaction. It includes soft radiation from the primary hard partons, secondary multiple parton interactions (MPI), and factorization-violating effects.  The invariant mass spectrum of the leading jet in $Z+$jet and $H+$jet events is directly sensitive to these effects, and we use a QCD factorization theorem to predict its dependence on the jet radius $R$, jet $p_T$, jet rapidity, and partonic process for both the perturbative and nonperturbative components of primary soft radiation.  We prove that the nonperturbative contributions involve only odd powers of $R$, and the linear $R$ term is universal for quark and gluon jets.  The hadronization model in \Pythiaeight agrees well with these properties.  The perturbative soft initial state radiation (ISR) has a contribution that depends on the jet area in the same way as the underlying event, but this degeneracy is broken by dependence on the jet $p_T$.  The size of this soft ISR contribution is proportional to the color state of the initial partons, yielding the same positive contribution for $gg\to Hg$ and $gq\to Zq$, but a negative interference contribution for $q\bar q\to Z g$. Hence, measuring these dependencies allows one to separate hadronization, soft ISR, and MPI contributions in the data.

\end{abstract}
%%%%%%%%%%%%%%%%%%%%%%%%%%%%%%%%%%%%%%%%%%%%%%%%%%%%%%%%%%%%%%%%%%%%%%%%%%%%%%%%

\maketitle

Soft hadronic activity plays a role in practically all but the most inclusive measurements at the LHC.
It is often an important yet hard-to-quantify source of uncertainty, so improving its theoretical understanding is vital.  One can consider four conceptually different sources for the effects that are experimentally associated with soft hadronic activity and the underlying event (UE):
%%%
\begin{enumerate}
  \item Perturbative soft radiation from the primary incoming and outgoing hard partons within factorization
  \item Nonperturbative soft effects within factorization associated with hadronization
  \item Multiple parton interactions (MPI)  at lower scales in the same proton-proton collision
  \item Factorization breaking contributions 
\end{enumerate}
%%%
For any given observable, the question is how much of each of these sources is required to describe the data. For example, it is known that including higher-order perturbative corrections (source 1) in parton-shower Monte Carlo programs can give a nontrivial contribution to traditional UE measurements~\cite{Cacciari:2009dp, Chatrchyan:2012tb}.

Traditionally, the UE activity is measured in regions of phase space away from hard jets~\cite{Affolder:2001xt, Acosta:2004wqa, Field:2005sa, Aaltonen:2010rm, Field:2010bc, ATLAS:2010osr, Aad:2010fh, ATLAS:2011zja, Aad:2011qe, Chatrchyan:2011id, Chatrchyan:2012tb}. These results are used to tune the MPI models which describe the UE in Monte Carlo programs~\cite{Buckley:2009bj, Skands:2010ak, Buckley:2011ms, ATLAS:2012uec, Gieseke:2012ft, Seymour:2013qka}. These models are then extrapolated into the jet region, where they are used to describe various jet observables, including the jet mass spectrum in dijet and Drell-Yan events~\cite{ATLAS:2012am, CMSjetmass}, which is an important benchmark jet observable at the LHC.

In this Letter, we directly consider the jet region and give a field-theoretic description of primary soft effects (sources 1 and 2), and discuss how to distinguish sources 1, 2, and 3. This is done 
using the dependence of the jet mass spectrum and its first moment on the jet radius $R$, jet momentum $p_T^J$, jet rapidity $y_J$, and participating partons.
We will not consider factorization-breaking effects here (see e.g. Ref.~\cite{Forshaw:2012bi}).

%%%%%%%%%%%%%%%%%%%%%%%%%%%%%%%%%%%%%%%%%%%%%%%%%%%%%%%%%%%%%%%%%%%%%%%%%%%%%%%%

We consider the jet mass spectrum in exclusive $pp\to Z+$1-jet and $pp\to H+$1-jet events. The factorization formula for $m_J \ll p_T^J$ that includes sources 1 and 2 is given by~\cite{Stewart:2010tn, Jouttenus:2011wh, Jouttenus:2013hs}
%%%
\begin{align}  \label{eq:fact}
  \frac{\df\sigma}{\df m_J^2 \df\Phi_2} 
 &= \sum_{\kappa,a,b} H_{\kappa}(\Phi_2) \!\int\!\! \df k_S \df k_B\, ({\cal I}_{\kappa_a a} {\cal I}_{\kappa_b b}\otimes\! f_a f_b)(k_B)
   \nn\\
  \times  J_{\kappa_J}&\!(m_J^2-2p_T^J k_S)\, S_\kappa (k_S , p^{\rm cut}\! - k_B,y_J,R)
\,.\end{align}
%%%
Here, $\Phi_2=\{p_T^J,y_J,Y\}$, $Y$ is the rapidity of the $Z/H+$jet system, $\kappa$ denotes the partonic channel, and $k_{S}$ and $k_B$ account for soft contributions to the jet mass $m_J^2$ and jet veto $p^{\rm cut}$ (which vetoes additional jets). The $H_\kappa(\Phi_2)$ contains the perturbative matrix elements for the hard process, and ${\cal I}_{\kappa_a a}{\cal I}_{\kappa_b b}\otimes f_a f_b$ describes perturbative collinear initial-state radiation convolved with the parton distribution functions. For the normalized jet mass spectrum, the dependence on $p^{\rm cut}$ largely drops out~\cite{Jouttenus:2013hs}. As a result, the shape of the jet mass spectrum is determined by the jet function $J_{\kappa_J}$, describing energetic final-state radiation, and by the soft function $S_\kappa$.    See also Refs.~\cite{Dasgupta:2012hg,Chien:2012ur}.

\begin{figure*}[t!]
\hfill%
\includegraphics[width=0.9\columnwidth]{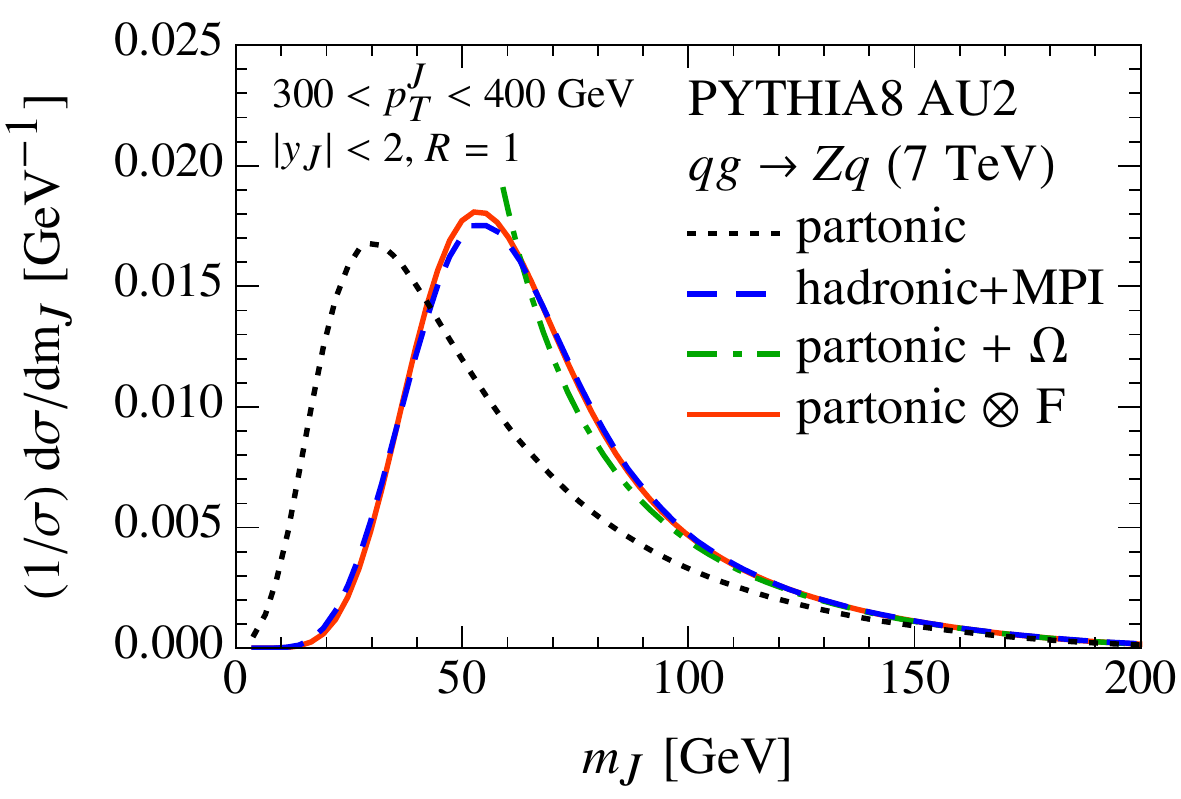}%
\hfill\hfill%
\includegraphics[width=0.9\columnwidth]{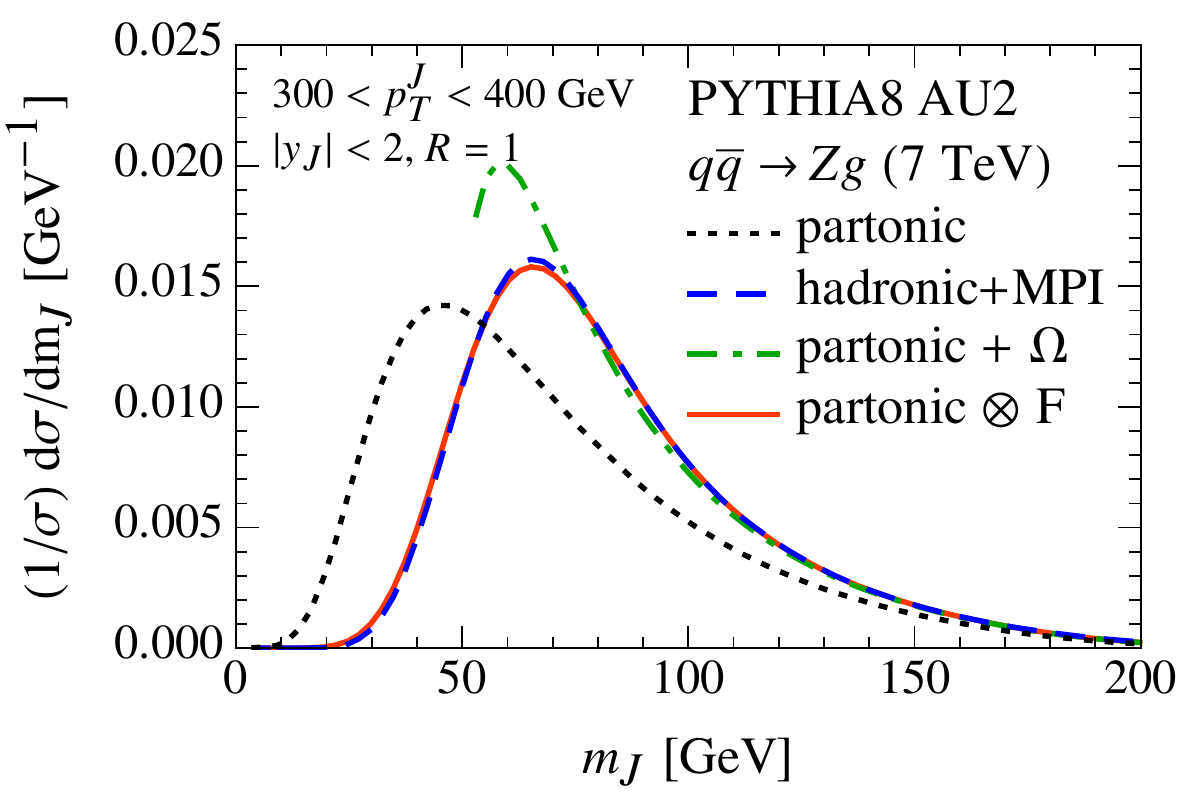}%
\hspace*{\fill}%
\vspace{-2ex}
\caption{For the jet mass spectrum in \Pythiaeight, the change from partonic to hadronization+MPI is described by a simple shift in the tail, and a simple convolution everywhere, for both quark jets (left panel) and gluon jets (right panel).}
\label{fig:pythiashift} 
\vspace{-1ex}
\end{figure*}

The soft function $S_\kappa$ describes the primary initial and final-state soft radiation. It depends on the jet through $y_J$ and $R$ but not $p_T^J$, and can be factorized as~\cite{Korchemsky:1999kt, Hoang:2007vb, Ligeti:2008ac}
%%%
\begin{align} \label{eq:Sfact}
S_\kappa(k_S, k_B,y_J, R)
&= \int\! \df k\, S_\kappa^{\rm pert}(k_S - k, k_B, y_J, R)
\\ \nn &\quad \times
F_\kappa(k, y_J, R) \bigl[1 + {\cal O}\big(\lqcd/k_B\big) \bigr]
,\end{align}
%%%
where $S_\kappa^{\rm pert}$ contains the perturbative soft contributions. $F_\kappa$ is a normalized nonperturbative shape function which encodes the smearing effect that the hadronization has on the soft momentum $k_S$. For $k_S\sim\lqcd$, the full $F_\kappa(k)$ is required and shifts the peak region of the jet mass spectrum to higher jet masses.

In the perturbative tail of the jet mass spectrum, where $k_S\gg \lqcd$, $S_\kappa$ can be expanded,
%%%
\begin{align} \label{eq:SOPE}
S_\kappa(k_S, y_J, R)
&= S_\kappa^\mathrm{pert}\bigl(k_S-\Omega_\kappa(R), y_J, R \bigr)
  \nn\\
 & \quad + \mathcal{O}\big({\lqcd^2}/{k_S^3}, \alpha_s\lqcd/k_S^2 \big)
\,,\end{align}
%%%
where $\Omega_\kappa(R) = \int\!\df k\, k\,F_\kappa(k) \sim \lqcd$ is a nonperturbative parameter. In this region factorization predicts a shift in the jet mass spectrum, which is described by $\Omega_\kappa(R)$. Below, we use the field-theoretic definition of $\Omega_\kappa$ to quantify its $R$ dependence and prove that it is independent of $y_J$. The above treatment provides an excellent description of hadronization in both $B$-meson decays and $e^+e^-$ event shapes~\cite{Bernlochner:2013bta, Abbate:2010xh}.

Factorization also underlies the Monte Carlo description of the primary collision, where $H$ corresponds to the hard matrix element, while $\cI$, $J$, and $S$ are described by parton showers, and $F$ corresponds to the hadronization models.  The standard parton shower paradigm does not completely capture interference effects between wide-angle soft emissions from different primary partons that appear at $\ord{\alpha_s}$ in $S_\kappa$. Monte Carlo programs include MPI (source 3), which are not in \eq{fact}. See Ref.~\cite{Gaunt:2014ska} for a recent discussion. For our numerical studies, we consider both \Pythiaeight~\cite{Sjostrand:2006za, Sjostrand:2007gs} with the ATLAS underlying event tune AU2-MSTW2008LO~\cite{ATLAS:2012uec} and \Herwig 2.7~\cite{Bahr:2008pv,Bellm:2013lba} with its default underlying event tune UE-EE-5-MRST~\cite{Seymour:2013qka}. Both give a reasonable description of the CMS jet mass spectrum in $Z$+jet events~\cite{CMSjetmass}. We also compare to the \Pythiaeight default tune 4C.

\begin{figure*}
\hfill%
\includegraphics[width=0.9\columnwidth]{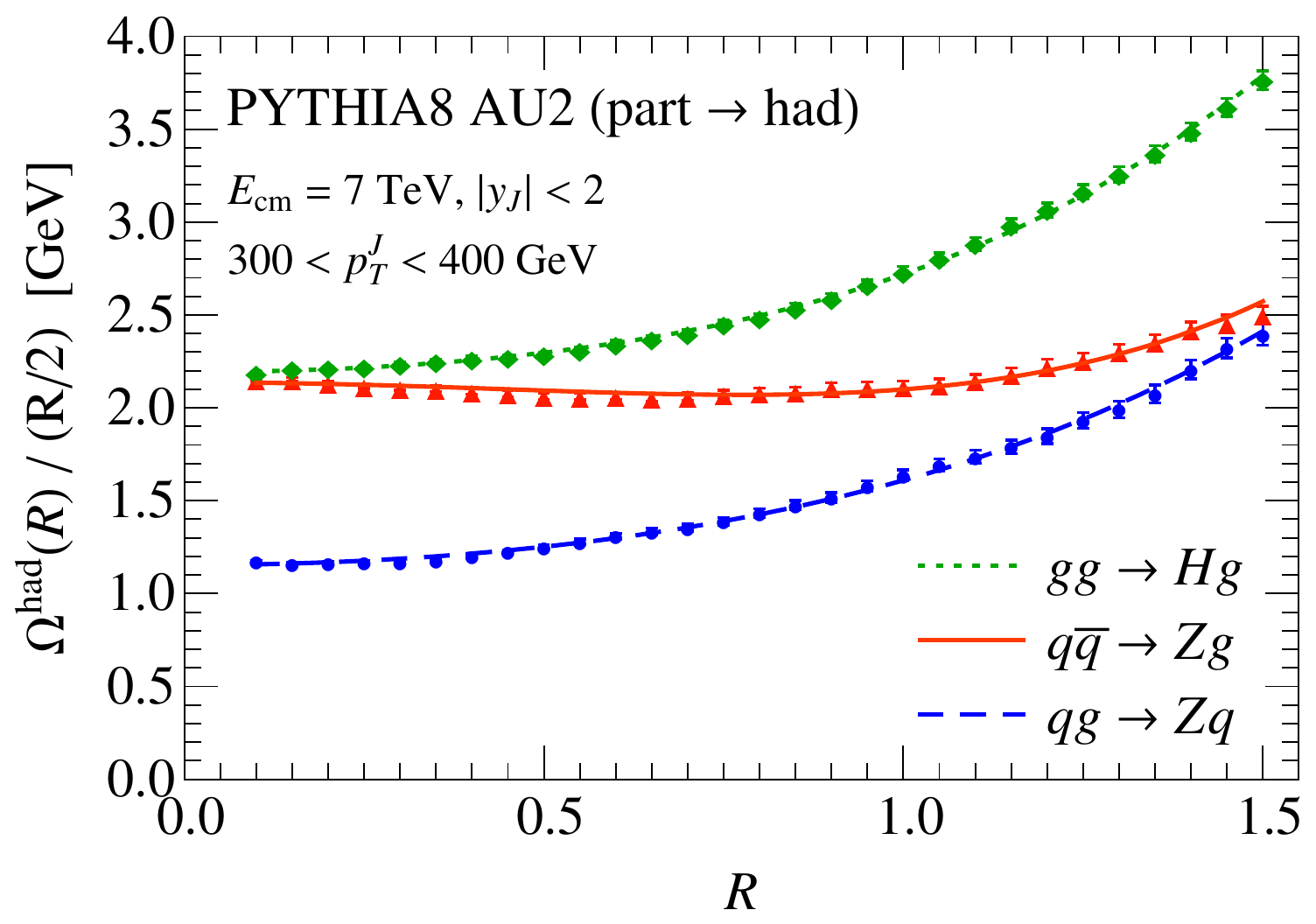}%
\hfill\hfill%
\includegraphics[width=0.9\columnwidth]{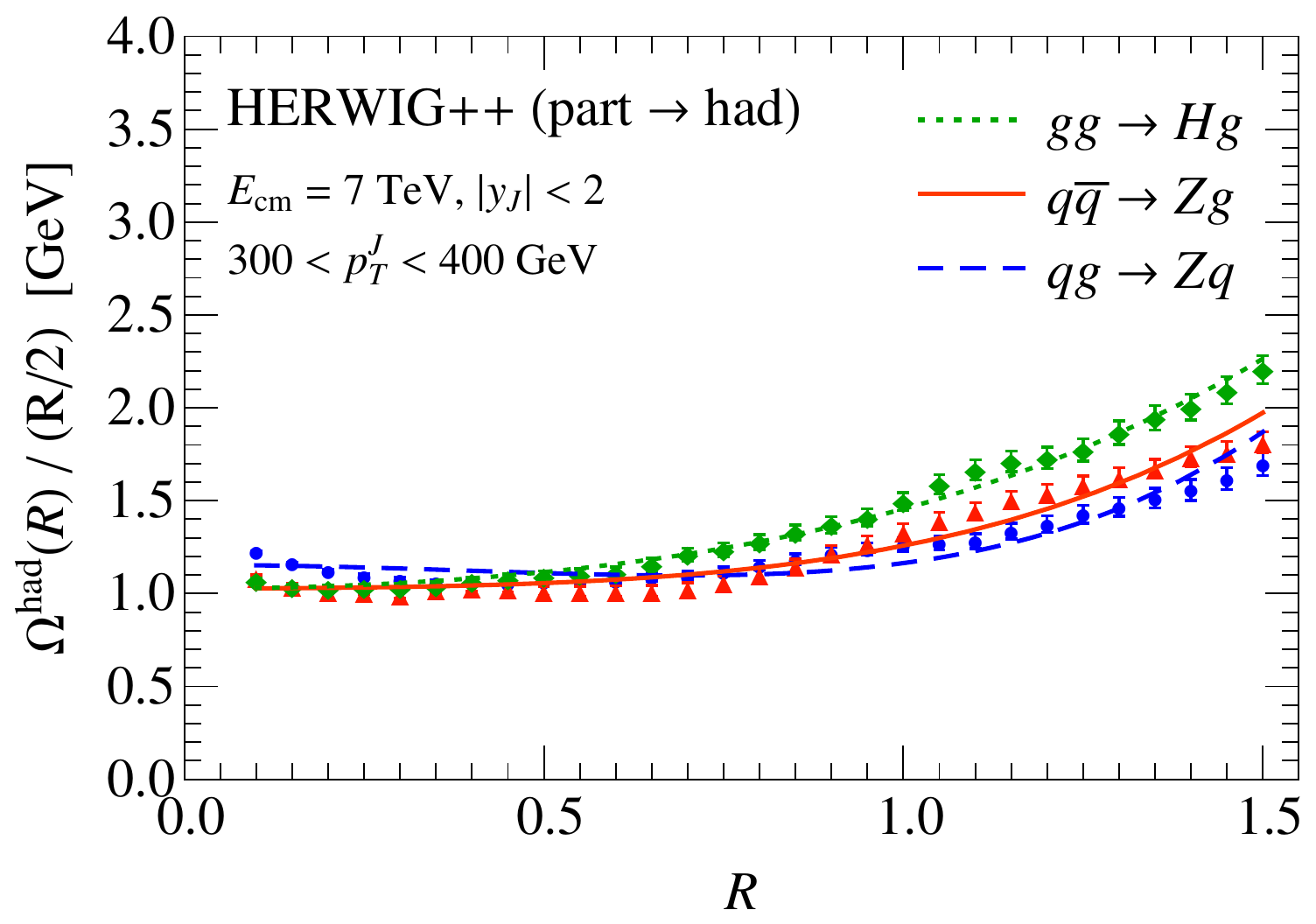}%
\hspace*{\fill}%
\vspace{-2ex}
\caption{The $R$ dependence of $\Omega_\kappa^{\rm had}(R)$ extracted from $M_1$ in \Pythiaeight (left panel) and \Herwig (right panel), shown as dots, triangles and squares for different channels. The fit using \eq{Omega_R} (shown by lines) demonstrates the agreement with factorization. The small-$R$ behavior only depends on whether the jet is initiated by a quark (blue dashed line) or gluon (orange solid and green dotted lines).}
\label{fig:Omegahad_R}
\vspace{-1ex}
\end{figure*}

We consider exclusive $Z/H+$jet events at $\Ecm = 7\,\TeV$ in both quark and gluon channels, with the leading jet within a certain range of $p_T^J$ and $y_J$, and we veto additional jets with $p_T^J > 50\,\GeV$.
The jets are defined using anti-$k_T$~\cite{Cacciari:2008gp, Cacciari:2011ma}.  In \fig{pythiashift}, we show the jet mass spectrum for quark and gluon jets with $R = 1$ after parton showering (black dotted line) and including both hadronization and MPI (blue dashed line). Equation~\eqref{eq:SOPE} predicts that for $m_J^2 \gg \lqcd p_T^J$ the nonperturbative corrections shift the tail of the jet mass spectrum by
%%%
\begin{equation}
m_J^2 = (m_J^2)^{\rm pert} + 2 p_T^J\, \Omega_\kappa(R)
\,.\end{equation}
%%%
We can regard the partonic result from \Pythiaeight as the baseline purely perturbative result. Choosing $\Omega=2.4\,{\rm GeV}$ for $qg \to Zq$ and $\Omega=2.7\,{\rm GeV}$ for $q\bar q\to Zg$ yields the green dot-dashed curves in \fig{pythiashift}. We see that the effect of both hadronization and MPI in the tail is well captured by this shift.  For hadronization, Eqs.~(\ref{eq:fact},\ref{eq:Sfact}) predict a convolution with a nonperturbative function,
%%%
\begin{align} \label{eq:pythiaconv}
\frac{\df \sigma_\kappa}{\df m_J^2}
&= \int \!\df k \,  \frac{\df \sigma_\kappa^{\rm partonic}}{\df m_J^2}(m_J^2-2 p_T^J k) \ F_\kappa(k)
\,.\end{align}
%%%
With the above $\Omega$'s, this convolution gives the red solid curves in \fig{pythiashift}, yielding excellent agreement with the hadronization+MPI result over the full range of the jet mass spectrum.\footnote{Here, $F_\kappa(k) = (4k/\Omega_\kappa^2)\, e^{-2k/\Omega_\kappa}$; the simplest ansatz  that satisfies the required properties: normalization, vanishing at $k=0$, falling off exponentially for $k\to \infty$, and having a first moment $\Omega_\kappa$. Fixing the value of $\Omega_\kappa$ from the tail, we find similar levels of agreement across all values of $p_T^J$, $y_J$, $R$, for all partonic channels, and for different jet veto cuts (including no jet veto).} Both hadronization and MPI populate the jet region with a smooth background of soft particles, which can explain why the MPI effect is reproduced alongside the hadronization by a convolution of the form \eq{pythiaconv}. This apparent degeneracy motivates us to determine the calculable behavior of the jet mass spectrum due to primary perturbative and nonperturbative soft radiation within factorization, study its dependence on $p_T^J$, $y_J$, and $R$, and compare these results to Monte Carlo program contributions for soft ISR, hadronization, and MPI.

We consider the first moment in $m_J^2$,
%%%
\begin{equation}
M_1 = \frac{1}{\sigma} \int\!\df m_J^2\, m_J^2\, \frac{\df\sigma}{\df m_J^2}
\,,\end{equation}
%%%
which tracks the shift observed in \fig{pythiashift}. 
Taking the first moment of \eq{fact} combined with \eqs{Sfact}{SOPE}, we can compute the dependence of primary soft radiation on $p_T^J$, $y_J$, $R$, and partonic channel, giving
%%%
\begin{align} \label{eq:M1primarysoft}
   M_1 = M_{1\,\kappa}^{\rm pert}(p_T^J,y_J,R) + 2 p_T^J\, \Omega_\kappa(R)
\,.\end{align}
%%%
Here, $M_{1\,\kappa}^{\rm pert}(p_T^J,y_J,R)$ contains all perturbative contributions, while $\Omega_\kappa(R)$ encodes the shift due to nonperturbative effects.

\begin{figure*}[t!]
\hfill%
\includegraphics[width=0.9\columnwidth]{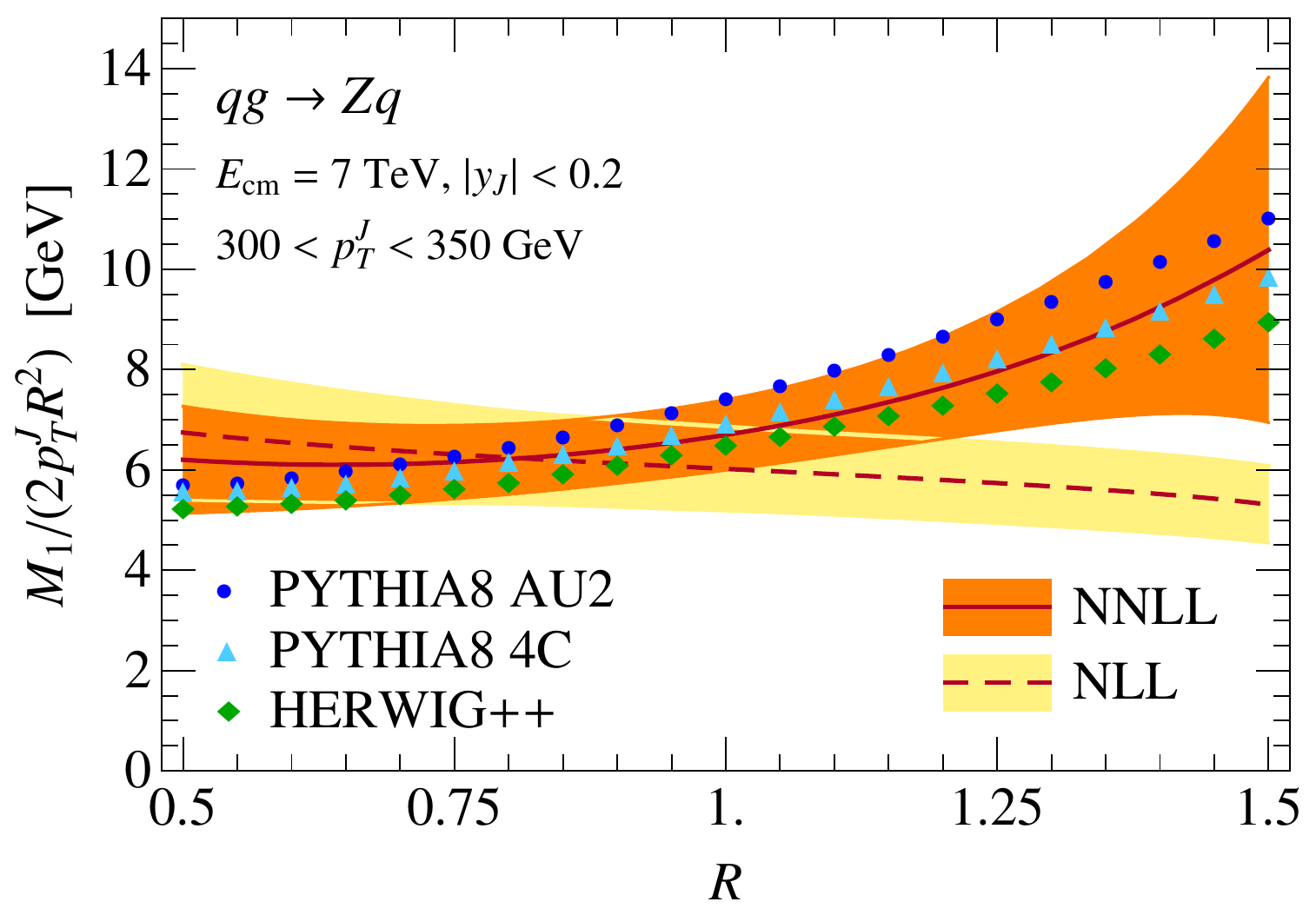}%
\hfill\hfill%
\includegraphics[width=0.9\columnwidth]{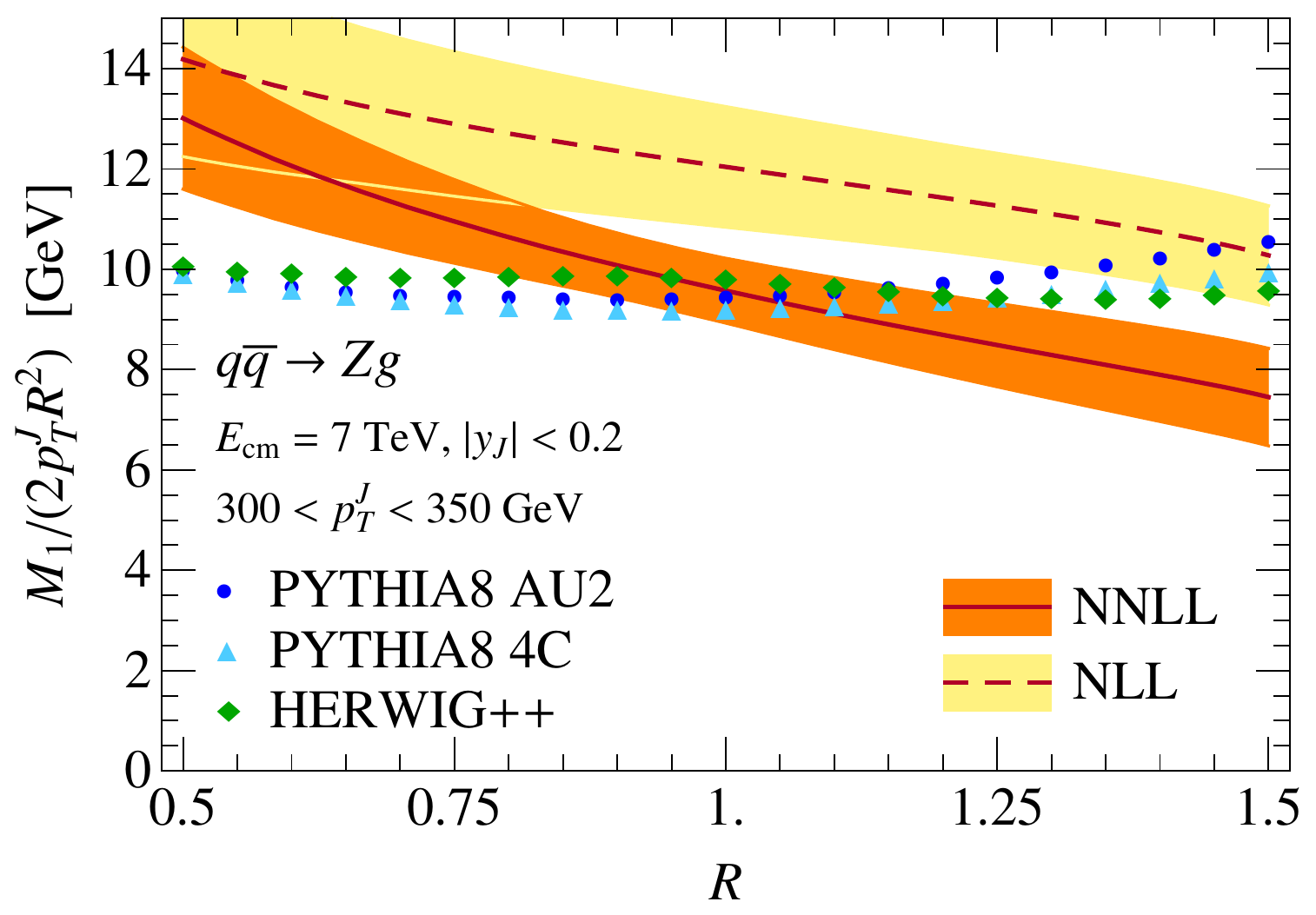}%
\hspace*{\fill}
\vspace{-2ex}
\caption{$R$ dependence of the perturbative jet mass moment $M_{1\,\kappa}^{\rm pert}$ at NLL and NNLL and the partonic jet mass moment $M_{1\,\kappa}^{\rm partonic}$ in \Pythiaeight (tune AU2 and 4C) and \Herwig for $qg\to Zq$ (left panel) and $q\bar q\to Zg$ (right panel). The soft ISR contribution $\sim R^4$ is well modeled by Monte Carlo programs for $qg\to Zq$, but not for the destructive interference in $q\bar q\to Zg$.}
\label{fig:M1R}
\vspace{-1ex}
\end{figure*}

\begin{figure}[t!]
\hfill%
\includegraphics[width=0.9\columnwidth]{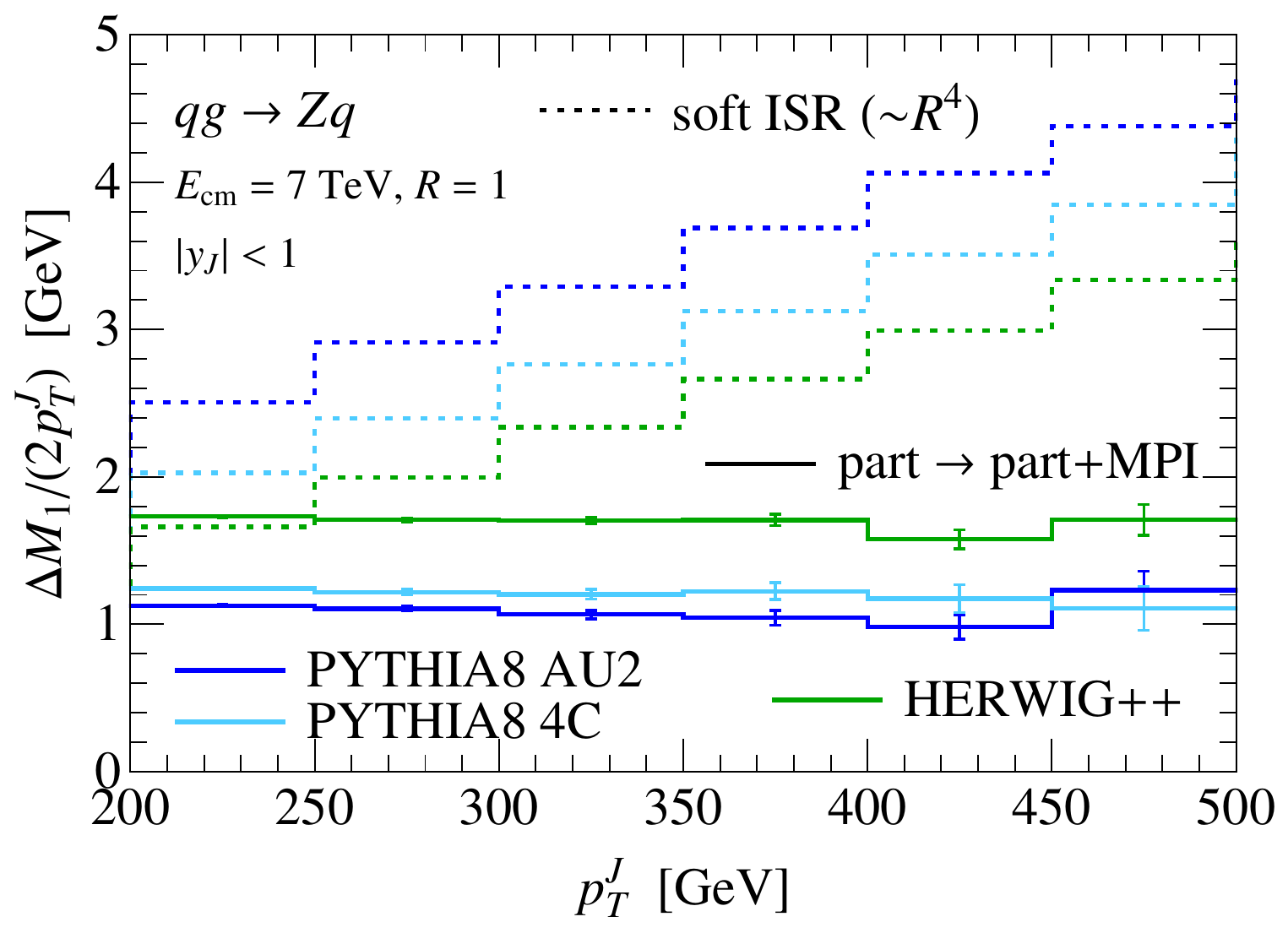}%
\hspace*{\fill}
\vspace{-2ex}
\caption{$p_T^J$ dependence of the $\sim R^4$ contributions to the jet mass moment in \Pythiaeight and \Herwig from MPI (solid lines from $\Upsilon^\mathrm{MPI}$), and soft ISR [dashed lines from $c_4^\kappa$ in \eq{M1fitform}] for $qg\to Zq$. They can be distinguished by their $p_T^J$ dependence.
}
\label{fig:DelM1MPI}
\vspace{-1ex}
\end{figure}

%%%%%%%%%%%%%%%%%%%%%%%%%%%%%%%%%%%%%%%%%%%%%%%%%%%%%%%%%%%%%%%%%%%%%%%%%%%%%%%%

For $pp\to H/Z+$jet, $\Omega_\kappa(R)$ is given by the vacuum matrix element of lightlike soft Wilson lines $Y_a$, $Y_b$, and $Y_J \equiv Y_J(y_J,\phi_J)$ along the beam and jet directions,
%%%
\begin{align} \label{eq:Omega}
\Omega_\kappa(R)
&= \int_0^1\! \df r \int_{-\infty}^{\infty}\! \df y \int_0^{2\pi}\! \df \phi\, f(r,y-y_J,\phi-\phi_J,R)
\nn \\ &\quad  \times\!
\bigl\langle 0 \bigl\lvert \bar T[Y_J^\dagger Y_b^\dagger Y_a^\dagger]\, \hat \cE_T(r,y,\phi) T[Y_a Y_b Y_J] \bigr\rvert 0 \bigr\rangle
.\end{align}
%%%
Here, the rapidity $y$, azimuthal angle $\phi$, and transverse velocity $r=p_T/m_T$ are measured with respect to the beam axis. The color representation of the Wilson lines depends on the partonic channel, giving the $\kappa$ dependence of $\Omega_\kappa$. The jet mass measurement function is
$f(r,y,\phi,R) = (\cosh y-r \cos \phi)\, \theta\bigl[ b(y,\phi,r) < R^2 \bigr]$
where $b(y,\phi,r)$ specifies the jet boundary. The matrix element involves the energy flow operator~\cite{Sveshnikov:1995vi, Cherzor:1997ak, Belitsky:2001ij, Lee:2006nr, Mateu:2012nk}
$\hat \cE_T(r,y,\phi) |X\rangle\! = \!\! \sum_{i \in X}\! m_{Ti} \de(r\!-\!r_i) \de(y\!-\!y_i)
 \de(\phi\!-\!\phi_i) |X\rangle$.
From \eq{Omega}, it follows immediately that $\Omega_\kappa(R)$ is independent of $p_T^J$. Using invariance under boosts and rotations, we can prove that it is also independent of $y_J$ and $\phi_J$~\cite{*[{See Supplemental Material at the end of this preprint}] [{}] supplement}.

Expanding \eq{Omega} for small $R$, we find~\cite{Marcantonini:2008qn, supplement}
%%%
\begin{align}\label{eq:Omega_R}
\Omega_\kappa(R)
&=  \frac{R}{2} \, \Omega_\kappa^{(1)} \!+\! \frac{R^3}{8}\, \Omega_\kappa^{(3)} \!+\! \frac{R^5}{32}\, \Omega_\kappa^{(5)} \!+\!  {\cal O}\Bigl[\Bigr(\frac{R}{2}\Bigl)^7\Bigr]
\,,\end{align}
%%%
where the $\Omega_\kappa^{(i)}$ are $R$ independent and only odd powers of $R$ occur.
This $R$ scaling of our nonperturbative operator for jet mass agrees with that found in Ref.~\cite{Dasgupta:2007wa} from a QCD hadronization model. Our operator definition implies a universality for the linear $R$ nonperturbative parameter in \eq{Omega_R}. For $R\to 0$ the beam Wilson lines fuse into a Wilson line in the conjugate representation to the jet, $Y_aY_b\to Y_{\bar J}$. The result is given by~\cite{supplement}
%%%
\begin{align}
\Omega_\kappa^{(1)}
= \int_0^1\! \df r'\,  \bigl\langle 0 \big| \bar T[Y_J^\dagger Y_{\bar J}^\dagger ]
 \hat \cE_\perp\!(r') T[Y_{\bar J} Y_J] \big|0 \bigr\rangle
\,,\end{align}
%%%
which only depends on whether the jet is a quark or gluon jet. For quarks, we can compare this to thrust in deep-inelastic scattering~\cite{Dasgupta:2002dc} where precisely this parameter $\Omega_q^{(1)}$ appears~\cite{Kang:2013nha}.

Consider next $M_{1\,\kappa}^{\rm pert}$ in \eq{M1primarysoft}. Dimensional analysis and the kinematical bound $m_J \lesssim p_T^J R$ imply that $M_{1\,\kappa}^{\rm pert}$ scales like $(p_T^J R)^2$. Resummation modifies the leading $R$ dependence to $R^{2-\gamma_\kappa}$, where $\gamma_\kappa  \sim \al_s > 0$. The soft function contains a contribution due to interference between ISR from the two beams~\cite{supplement},
%%%
\begin{align} \label{eq:soft_interf_short}
S^{\rm pert}_\kappa(k_S) \supset \frac{\alpha_s C_\kappa}{\pi}\, R^2\,\frac{1}{\mu} \Bigl(\frac{\mu}{k_S}\Bigr)_+
\,.\end{align}
%%%
The extra $R^2$ for soft ISR causes it to contribute to $M_{1\,\kappa}^{\rm pert}$ as $(p_T^J)^2 R^4$ with the color factors
%%%
\begin{align} \label{eq:ISRcolor}
C_{qg\to q} = C_{gg\to g} &= \frac{C_A}{2} = \frac{3}{2}
\,, \nn \\
C_{q\bar q\to g} &= C_F - \frac{C_A}{2} = - \frac{1}{6}
\,.\end{align}
%%%

The above factorization results can be compared to \Pythiaeight and \Herwig, where we find that the dependence of $M_1$ on $p_T^J$, $y_J$, $\kappa$, 
is well described by
%%%
\begin{align}  \label{eq:M1mc}
M_1 &=  M_{1\,\kappa}^{\rm partonic}(p_T^J,y_J,R) + 2 p_T^J\, \Omega^{\rm had}_\kappa(R)
\nn \\ & \quad
+ 2 p_T^J\, \Bigl[ \Upsilon^{\rm MPI}(y_J,R) +\Omega^{\rm MPI}_\kappa(y_J,R) \Bigr]
\,.\end{align}
%%%
Here, $M_{1\,\kappa}^{\rm partonic}$ is the partonic contribution, $\Omega_\kappa^{\rm had}$ is defined by partonic $\to$ hadronic, and $\Upsilon^{\rm MPI}$ by partonic $\to$ partonic+MPI. The small remainder from hadronization of the MPI, $\Omega^{\rm MPI}_\kappa$,  is defined to ensure the sum of terms yields the full partonic $\to$ hadronic+MPI.   Note that hadronization and MPI contributions are each individually described by shifts to $M_1$. Also, the independence of $\Omega_\kappa$ to $y_J$ and $\phi_J$ is observed in both \Pythiaeight and \Herwig~\cite{supplement}. Equation~\eqref{eq:M1mc} contains MPI contributions with no analog in \eq{M1primarysoft}.

The hadronization $\Omega^{\rm had}_\kappa(R)/(R/2)$ from \Pythiaeight and \Herwig is shown in \fig{Omegahad_R} for different channels. For $R \ll 1$, $\Omega^{\rm had}_\kappa(R)$ is linear in $R$ and has the same slope for the two channels involving gluon jets, as predicted by factorization. For \Pythiaeight, all channels differ for large $R$ and can be fit to the factorization form in \eq{Omega_R}. For the quark jet we extract $\Omega_q^{(1)} = 1.2\,\GeV$ and for gluon jets $\Omega_g^{(1)}=2.2\,\GeV$. For $qg\to Zq$ and $gg\to Hg$ the $R$ dependence is strong enough that an additional $R^2$ contribution is disfavored in the fit. For \Herwig, the dependence on higher powers of $R$ is much weaker, and $\Omega_g^{(1)} \approx \Omega_q^{(1)}$. The full set of fit coefficients is in \cite{supplement}.

%%%%%%%%%%%%%%%%%%%%%%%%%%%%%%%%%%%%%%%%%%%%%%%%%%%%%%%%%%%%%%%%%%%%%%%%%%%%%%%%

In \fig{M1R}, we compare our perturbative next-to-leading logarithmic (NLL) and next-to-next-to-leading logarithmic (NNLL) factorization predictions~\cite{Jouttenus:2013hs} for $M_{1\,\kappa}^{\rm pert}$ to the corresponding $M_{1\,\kappa}^{\rm partonic}$ from \Pythiaeight and \Herwig as a function of $R$, dividing by the leading $R^2$ dependence. The $R^4$ contribution from soft ISR only enters at NNLL and is seen in the rise at large $R$ for $qg \to Zq$ (left panel). This effect is partially modeled by soft emissions in the parton shower, which explains the similar $R^4$ contribution for $qg \to Zq$ in \Pythiaeight and \Herwig. For $q\bar q \to Zg$ (right panel) \eqs{soft_interf_short}{ISRcolor} predict the $R^4$  contribution from soft ISR to be negative, which we observe at NNLL. This negative interference effect is not captured by these Monte Carlo programs.

The apparent ambiguity between $R^4$ contributions from soft ISR and MPI can be resolved through their $p_T^J$ dependence.
In \fig{DelM1MPI}, we show the $R^4$ component $c_4^\kappa$ of the partonic moment, obtained by fitting
%%%
\begin{align}  \label{eq:M1fitform}
\frac{M_{1\,\kappa}^{\rm part}}{2p_T^J R^2} &=  c_2^\kappa\, R^{-\gamma_\kappa}  +  c_4^\kappa \, R^2
\,,\end{align}
%%%
and also the MPI contribution to the moment, $\Upsilon^\mathrm{MPI}/ R^2 \sim R^2$. The differences between various tunes for $c_4^\kappa$ and $\Upsilon^\mathrm{MPI}$ reflects their apparent ambiguity, whereas their 
sum agrees much better. The $p_T^J$ dependence clearly resolves the ambiguity: $c_4^\kappa \sim p_T^J$ as predicted by factorization, whereas $\Upsilon^\mathrm{MPI}$ is independent of $p_T^J$. As shown in Ref.~\cite{supplement}: the channel dependence could also be used to separate soft ISR from MPI: $c_4^\kappa$ depends on the color channel as in \eq{ISRcolor}, whereas $\Upsilon^\mathrm{MPI}$ is channel independent. Also, the $y_J$ dependence of soft ISR is quite different between \Herwig and \Pythiaeight.

To conclude, we have used QCD factorization to predict the properties of the perturbative and nonperturbative components of primary soft radiation for jet mass in $pp\to H/Z+$jet. We have shown that the nonperturbative soft effects involve odd powers of $R$ and are universal for quark and gluon jets for $R\ll 1$. Hadronization models in Monte Carlo programs agree with these predictions. The perturbative soft radiation has a contribution that scales like $R^4$, just like the contribution from MPI. These components depend differently on $p_T^J$ and on the partonic process. Hence, separately measuring quark and gluon channels in Drell-Yan events and in different bins of $p_T^J$ provides the possibility to clearly distinguish between MPI and primary soft radiation.

%%%%%%%%%%%%%%%%%%%%%%%%%%%%%%%%%%%%%%%%%%%%%

We thank Jesse Thaler and Simon Pl\"atzer for helpful conversations.
This work was supported in part by the Office of Nuclear Physics of the U.S.
Department of Energy under Grant No.  DE-SC0011090,
the DFG Emmy-Noether Grant No. TA 867/1-1, and the Marie Curie International Incoming Fellowship PIIF-GA-2012-328913 within the 7th European Community Framework Program. We thank the Erwin Schr\"odinger Institute for hospitality while portions of this work were completed.

%%%%%%%%%%%%%%%%%%%%%%%%%%%%%%%%%%%%%%%%%%%%%

\bibliography{../pp}

\setcounter{equation}{0}
\renewcommand{\theequation}{S-\arabic{equation}}

\clearpage

%%%%%%%%%%%%%%%%%%%%%%%%%%%%%%%%%%%%%%%%%%%%%%%%%%%%%%%%%%%%%%%%%%%%%%%%%%%%%%%%
\section*{Supplemental Material}
%%%%%%%%%%%%%%%%%%%%%%%%%%%%%%%%%%%%%%%%%%%%%%%%%%%%%%%%%%%%%%%%%%%%%%%%%%%%%%%%

%===============================================================================
\subsection*{Nonperturbative corrections}
%===============================================================================

\begin{figure*}[t!]
\includegraphics[width=0.8\textwidth]{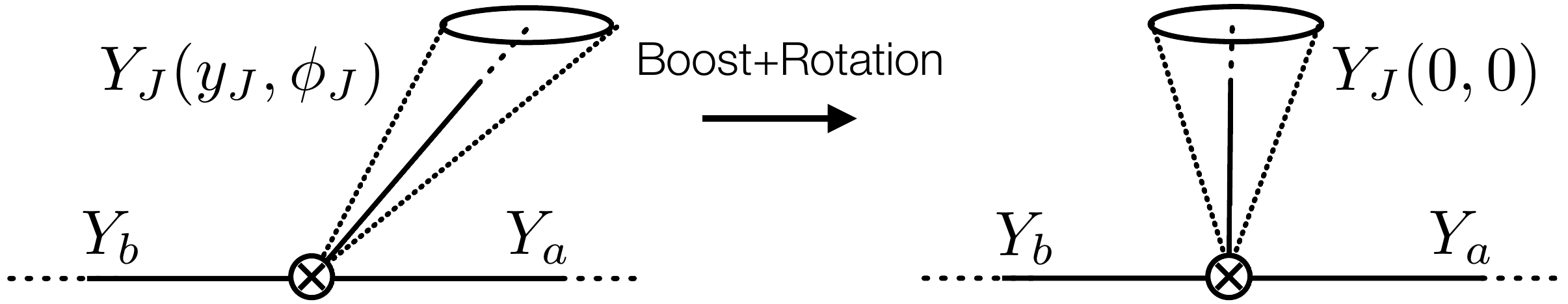}
\caption{Boost by $-y_J$ along the beam direction and rotation by $-\phi_J$ around the beam direction used to show that $\Omega_\kappa$ is independent of $y_J$ and $\phi_J$.}
\label{fig:yJboost}
\end{figure*}

\begin{figure*}[t!]
\hfill%
\includegraphics[width=0.9\columnwidth]{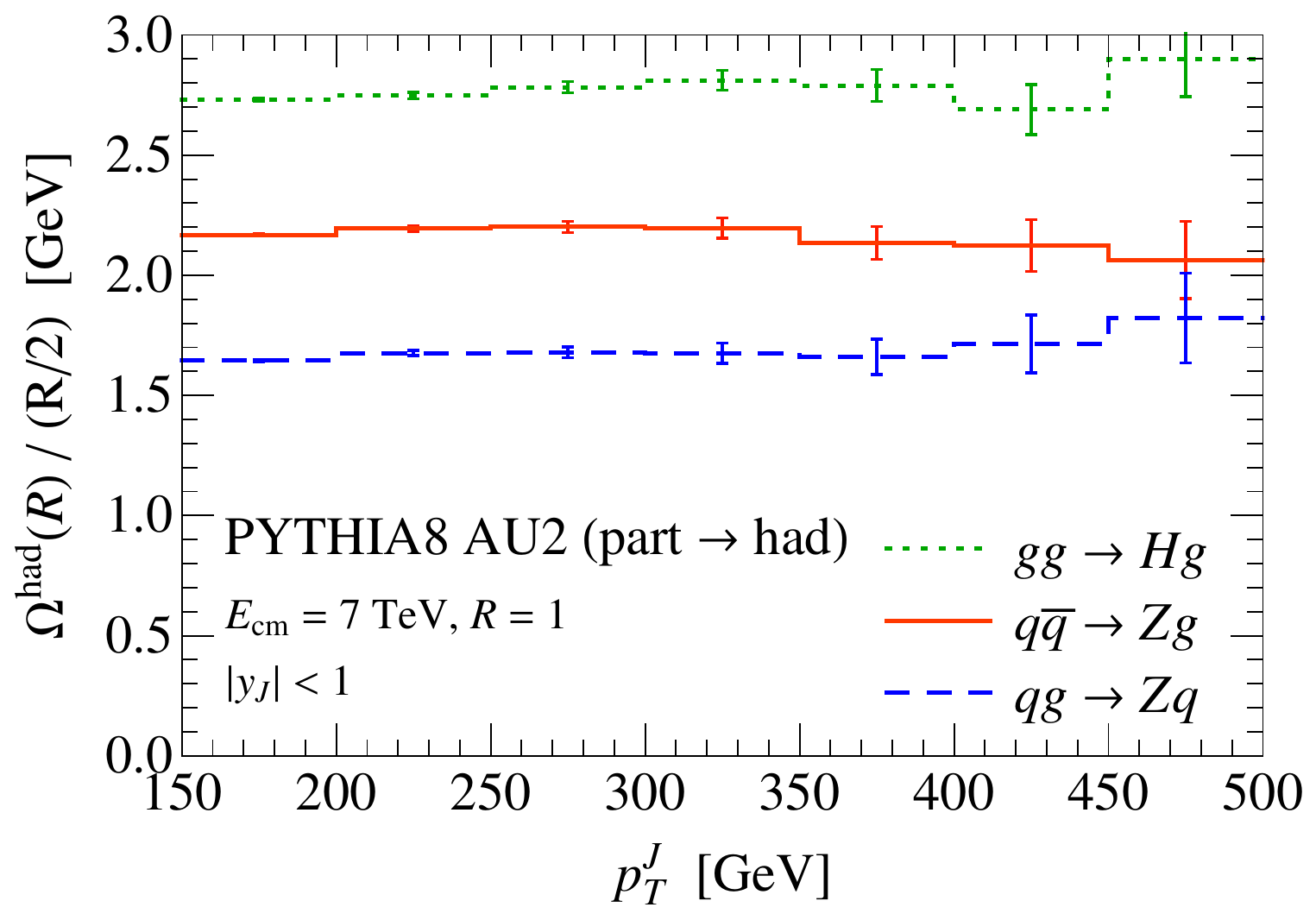}%
\hfill\hfill%
\includegraphics[width=0.9\columnwidth]{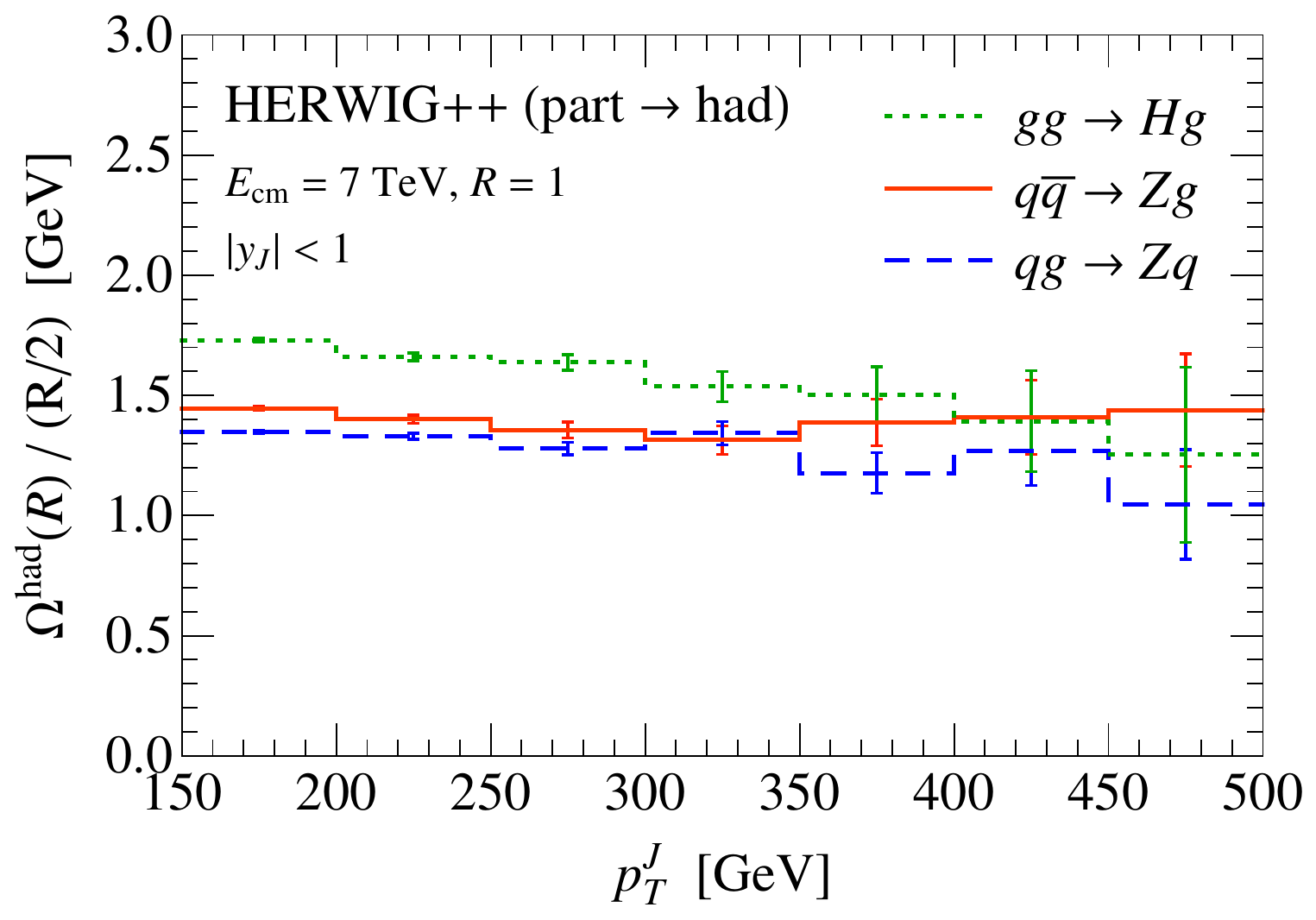}%
\hspace*{\fill}
\vspace{-2ex}
\caption{$p_T^J$ dependence of $\Omega_\kappa^{\rm had}(R)$ for \Pythiaeight (left panel) and \Herwig (right panel).}
\label{fig:Omegahad_pTJ}
\vspace{-1ex}
\end{figure*}

\begin{figure*}[t!]
\hfill%
\includegraphics[width=0.9\columnwidth]{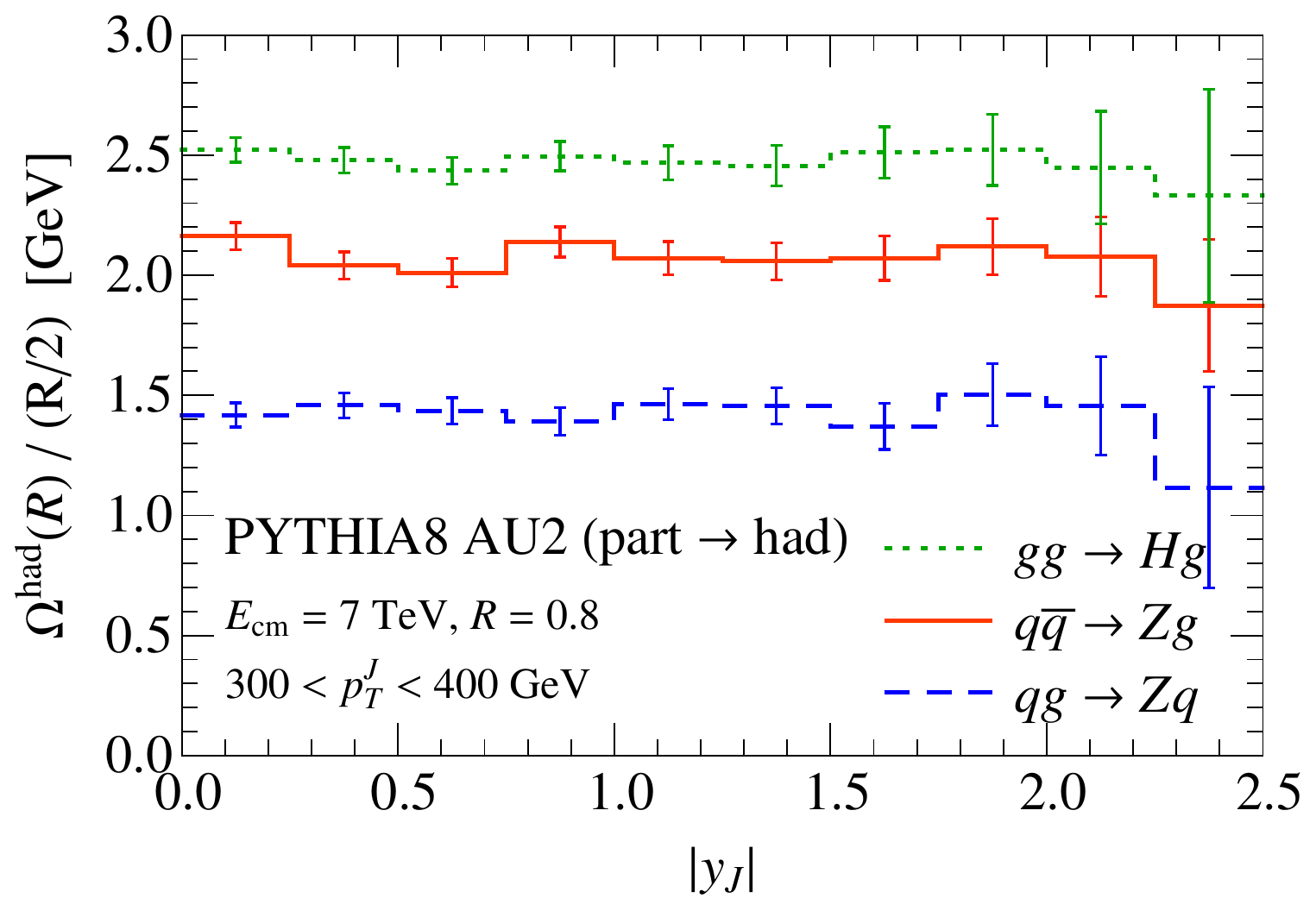}%
\hfill\hfill%
\includegraphics[width=0.9\columnwidth]{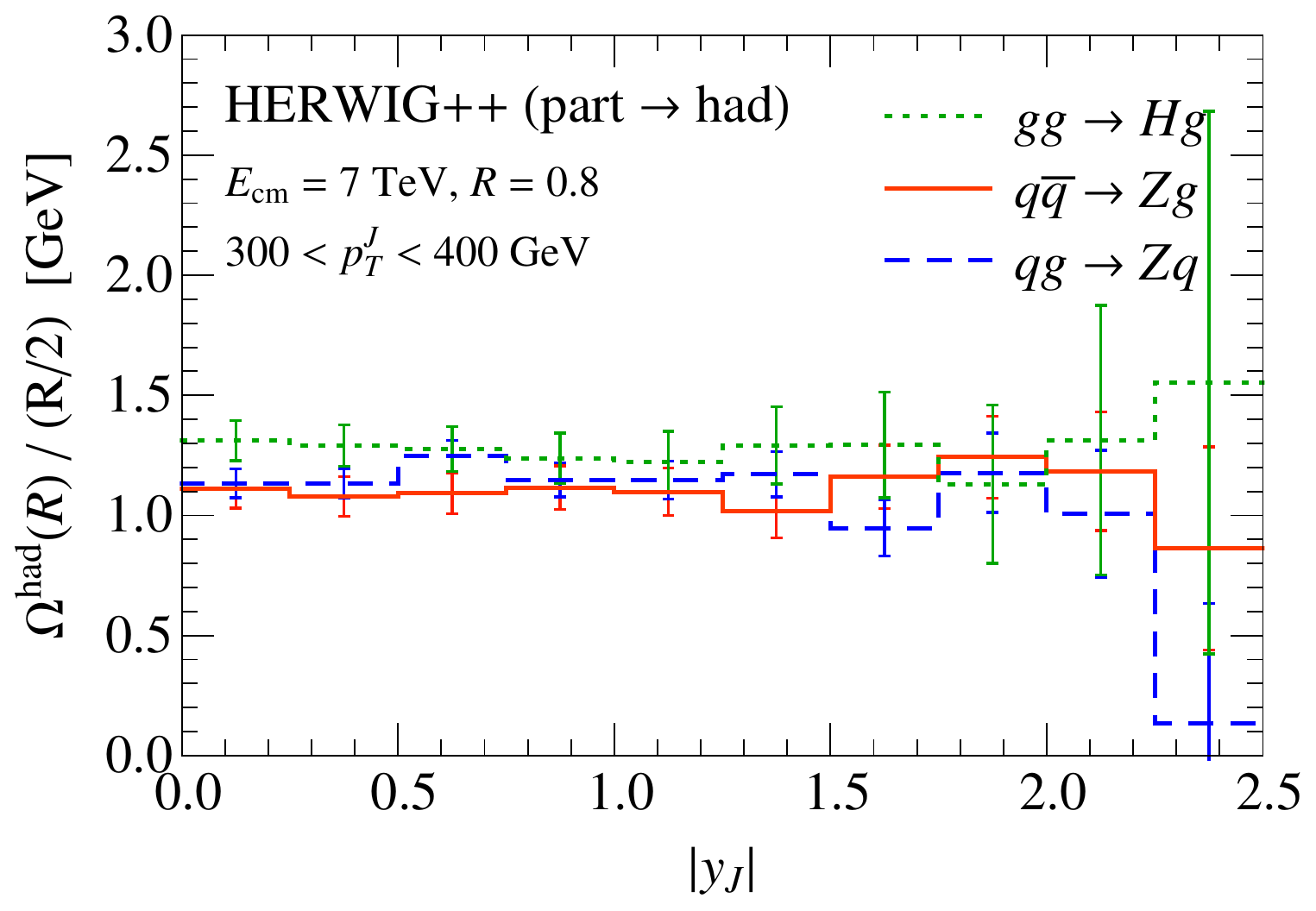}%
\hspace*{\fill}
\vspace{-2ex}
\caption{Jet rapidity dependence of $\Omega_\kappa^{\rm had}(R)$ for \Pythiaeight (left panel) and \Herwig (right panel).}
\label{fig:Omegahad_yJ}
\vspace{-1ex}
\end{figure*}

The leading hadronization effects in the first jet mass moment and in the tail of the jet mass spectrum are described by the parameter $\Omega_\kappa(R)$, which is defined by
%%%
\begin{align} \label{eq:Omega2}
\Omega_\kappa(R)
&= \int_0^1\! \df r \int_{-\infty}^{\infty}\! \df y \int_0^{2\pi}\! \df \phi\, f(r,y-y_J,\phi-\phi_J,R)
\nn \\ &\quad  \times\!
\bigl\langle 0 \bigl\lvert \bar T[Y_J^\dagger Y_b^\dagger Y_a^\dagger]\, \hat \cE_T(r,y,\phi) T[Y_a Y_b Y_J] \bigr\rvert 0 \bigr\rangle
.\end{align}
%%%
Here, $Y_a$ and $Y_b$ are incoming soft Wilson lines along the beam directions in the color representation of the incoming primary hard partons. $Y_J \equiv Y_J(y_J, \phi_J)$ is an outgoing soft Wilson line along the jet direction in the color representation of the outgoing hard parton. The color contractions between the Wilson lines are suppressed in \eq{Omega2}, but are normalized such that
$\langle 0 | Y_J^\dagger Y_b^\dagger Y_a^\dagger Y_a Y_b Y_J |0 \rangle = 1$.  In \fig{Omegahad_pTJ} we show that $\Omega_\kappa^{\rm had}(R)$ in \Pythiaeight and \Herwig is independent of $p_T^J$ over the large range of $p_T^J$ considered. (The $qg\to Zq$ and $gg\to Hg$ channels in \Herwig have a small downward trend in $p_T^J$.) 

For $e^+e^-\to$ dijets, a boost allows one to derive important universality properties of the corresponding $\Omega$ parameter~\cite{Lee:2006nr}. Here for \eq{Omega2}, boosting by $-y_J$ along the beam axis and rotating by $-\phi_J$, as illustrated in \fig{yJboost}, the Wilson lines and energy flow operator transform as
%%%
\begin{align}
Y_{a,b} &\to Y_{a,b}
\,, \nn \\
Y_J(y_J,\phi_J) &\to Y_J(0,0)
\,, \nn \\
 \hat \cE_T(r,y,\phi) &\to  \hat \cE_T(r,y-y_J,\phi-\phi_J)
\,.\end{align}
%%%
Changing variables $y\to y+y_J$ and $\phi\to \phi+\phi_J$ then yields an expression depending only on $y$ and $\phi$, which thus shows that $\Omega_\kappa(R)$ in \eq{Omega2} is independent of $y_J$ and $\phi_J$. We therefore set $y_J=\phi_J=0$ in the following. Note that unlike for $e^+e^-\to$ dijets, the matrix element is not independent of $y$ and $\phi$, so these dependencies in the measurement $f(r,y,\phi)$ do not generically decouple. In \fig{Omegahad_yJ} we show that $\Omega_\kappa^{\rm had}$ obtained from Monte Carlo programs does not depend on the jet rapidity $y_J$. The behaviour with $p_T^J$ and $y_J$ shown in these plots does not depend on the value of $R$. In Figs.~\ref{fig:Omegahad_pTJ} and \ref{fig:Omegahad_yJ} we see again that in \Pythiaeight the overall size of $\Omega_\kappa^{\rm had}(R)$ depends on the channel, being larger for the channels with a gluon jet. In contrast, $\Omega_\kappa^{\rm had}(R)$ in \Herwig is smaller and of similar sizes for all channels.

\begin{figure*}[t!]
\includegraphics[width=0.8\textwidth]{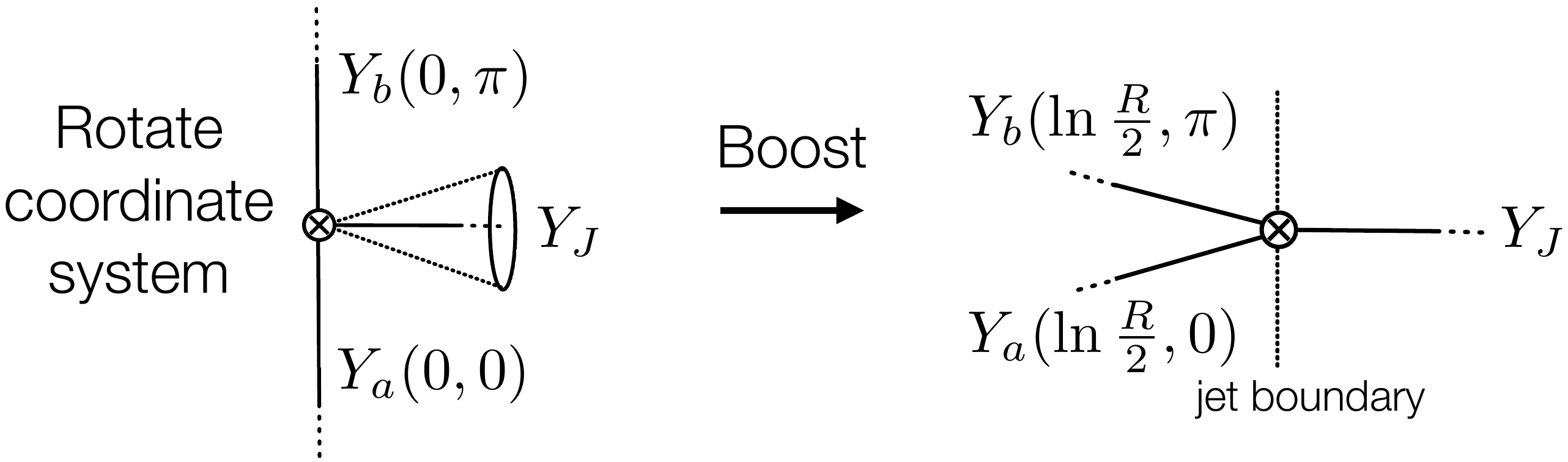}
\caption{Changing coordinates and boosting by $\ln(R/2)$ along the jet direction in order to expand around small $R$.}
\label{fig:Rexpansion}
\end{figure*}

To discuss the $R$ dependence of $\Omega_\kappa(R)$, we switch to coordinates $\{y', \phi', r'\}$ measured with respect to the jet axis. This gives
%%%
\begin{align} \label{eq:Omegajetaxis}
\Omega_\kappa(R)
&= \int_0^1\! \df r' \int_{-\infty}^{\infty} \! \df y' \int_0^{2\pi}\! \df \phi' \,
f(r',\!y',\!\phi',\!R)
\nn \\ & \quad \times
\bigl\langle 0 \bigl\lvert \bar T[Y_J^\dagger Y_b^\dagger(0, \pi)  Y_a^\dagger(0, 0)] \,\hat \cE_\perp(r', y', \phi')\,
\nn \\ & \qquad \times
 T[Y_a(0,0) Y_b(0,\pi) Y_J] \bigr\rvert  0 \bigr\rangle
\,,\end{align}
%%%
where the incoming beam Wilson lines $Y_{a,b}$ point in the $(y',\phi')=(0,0)$ and $(0,\pi)$ directions, and $r'=p_\perp/m_\perp$.
The measurement function in the original coordinates in \eq{Omega2} is given by
%%%
\begin{equation}
f(r,y,\phi,R) = (\cosh y-r \cos \phi)\, \theta\bigl[ b(y,\phi,r) < R^2 \bigr]
\,,\end{equation}
%%%
where we use $b(y,\phi,r)= 2(\cosh y- r \cos \phi)$ to define the jet boundary. In the primed coordinates it takes the form
%%%
\begin{align} \label{eq:fp}
&f(r',y',\phi',R)
\nn \\ & \qquad
= e^{-y'} \theta\Big[e^{2y'} - 2r'^2 \cos^2 \phi' + 1
\nn \\ & \qquad \quad
  - \frac{2}{R^2}\sqrt{4 + R^4(r'^4 \cos^4 \phi' - r'^2 \cos^2 \phi')}\Big]
\,.\end{align}
%%%
Boosting along the \emph{jet axis} by $\ln(R/2)$ as in \fig{Rexpansion}, the Wilson lines and energy flow operator transform as
%%%
\begin{align}
Y_a(0,0) &\to Y_a(\ln \tfrac{R}{2},0)
\,, \nn \\
Y_b(0,\pi) &\to Y_b(\ln \tfrac{R}{2},\pi)
\,, \nn \\
Y_J &\to Y_J
\,, \nn \\
\hat \cE_\perp(r',y',\phi') &\to  \hat \cE_\perp(r',y'\!+\!\ln \tfrac{R}{2},\phi')
\,.\end{align}
%%%
In these coordinates, the beam Wilson lines are an angle $\theta = 4\tan^{-1}(R/2) \simeq 2R$ apart.

We can now expand the result in $R$. To leading order in $R$, the measurement in \eq{fp} becomes
%%%
\begin{align}\label{eq:fR_expand}
f(r', y', \phi', R) = e^{-y'}  \theta\bigl[y' + \ln(R/2) \bigr] \bigl[ 1 +  \ord{R^2} \bigr]
\,.\end{align}
%%%
For the leading term in the $R\to 0$ limit, the beam Wilson lines fuse
%%%
\begin{align}\label{eq:YR_expand}
Y_a(\ln \tfrac{R}{2},0) Y_b(\ln \tfrac{R}{2},\pi) = Y_{\bar J}(-\infty,0) + {\cal O}(R^2)
\,,\end{align}
%%%
where $Y_{\bar J}$ is an incoming Wilson line along the direction opposite to the jet and in the appropriate conjugate color representation that forms a color singlet with the outgoing jet Wilson line $Y_J$. Since we now have two Wilson lines along the jet axis, we can boost along the jet axis to eliminate the $y'$ dependence. Integrating over $\phi'$ then yields
the result in \eq{Omega_R}, namely
%%%
\begin{align}\label{eq:Omega_R2}
\Omega_\kappa(R)
&=  \frac{R}{2} \, \Omega_\kappa^{(1)} \!+\! \frac{R^3}{8}\, \Omega_\kappa^{(3)} \!+\! \frac{R^5}{32}\, \Omega_\kappa^{(5)} \!+\!  {\cal O}\Bigl[\Bigr(\frac{R}{2}\Bigl)^7\Bigr]
\,,\end{align}
%%%
where the coefficient of the leading term comes from integrating the measurement function over $y'$,
%%%
\begin{align}
\int_{-\infty}^\infty\!  \df y' \,  e^{-y'}  \theta\bigl[y' + \ln(R/2) \bigr] = \frac{R}{2}
\,.\end{align}
%%%
The leading nonperturbative parameter in \eq{Omega_R2} is given by a universal matrix element
%%%
\begin{align}
\Omega_\kappa^{(1)}
= c_e \int_0^1\! \df r'  g_e(r')
  \bigl\langle 0 \big| \bar T[Y_J^\dagger Y_{\bar J}^\dagger ]
 \hat \cE_\perp\!(r') T[Y_{\bar J} Y_J] \big|0 \bigr\rangle
.\end{align}
%%%
It depends on the color representation of the Wilson line (quark vs.~gluon) but not the full original color configuration. To extend our result to a more general jet measurement $e$, we included the parameters $c_e$ and $g_e(r')$, which in our case simply are given by $c_e=g_e(r')=1$. In general $c_e$ is the calculable coefficient for the observable $e$~\cite{Lee:2006nr} obtained here by integrating over our $y'$ variable. The calculable function $g_e(r')$ encodes the dependence on hadron mass effects~\cite{Mateu:2012nk}.

\begin{table}
 \begin{tabular}{l|c|cccl}
 \hline\hline
   & $\kappa$ & $\Omega_\kappa^{(1)}$ & $\Omega_\kappa^{(3)}$ & $\Omega_\kappa^{(5)}$ & [GeV]\\
 \hline
  \Pythiaeight AU2 & $qg\to q$ & $1.2$ & $1.5$ & $1.3$ \\
  \Pythiaeight 4C & $qg\to q$ & $1.1$ & $0.7$ & $2.0$ \\
  \Herwig  & $qg\to q$ & $1.2$ & $-0.9$ & $4.0$ \\
  \hline
  \Pythiaeight AU2 & $q\bar q\to g$ & $2.1$ & $-0.9$ & $3.0$ \\
  \Pythiaeight 4C & $q\bar q\to g$ & $2.1$ & $-1.4$ & $3.4$ \\
  \Herwig  & $q\bar q\to g$ &  $1.0$ & $0.3$ & $2.4$ \\
  \hline
  \Pythiaeight AU2 & $gg\to g$ & $2.2$ & $1.5$ & $2.4$ \\
  \Pythiaeight 4C & $gg\to g$ & $2.1$ & $0.4$ & $3.0$ \\
  \Herwig  & $gg\to g$ &  $1.0$ & $1.3$ & $1.5$ \\
 \hline\hline
 \end{tabular}
 \caption{Fit coefficients for $\Omega_\kappa(R)$ in \eq{Omega_R} for different Monte Carlo programs and tunes which give the lines shown in Fig.~\ref{fig:Omegahad_R}.}
\label{tab:Omegakappa}
\end{table}

The expansions in \eqs{fR_expand}{YR_expand} can be carried out to higher orders in $R$, using Ref.~\cite{Marcantonini:2008qn} to expand the Wilson lines about the $\bar J$ direction, and lead to new nonperturbative matrix elements, collectively denoted as $\Omega_\kappa^{(3,5)}$ in \eq{Omega_R2}. Terms with an odd number of gauge field components that are transverse to the jet direction vanish due to parity invariance. Together with the overall factor of $R$, this implies that $\Omega_{\kappa}(R)$ only contains odd powers of $R$. The coefficients of the fits shown in \fig{Omegahad_R} are given in Table~\ref{tab:Omegakappa}. The leading coefficient in $R$, $\Omega_\kappa^{(1)}$, is the same for quark and gluon jets, while the higher coefficients are quite different for all three channels. The higher coefficients $\Omega_\kappa^{(3)}$ and $\Omega_\kappa^{(5)}$ strongly depend on the Monte Carlo program and tune. They are also correlated so their separation is not well constrained by the fit. The fact that all the coefficients are of similar size confirms that $R/2$ is indeed the appropriate expansion parameter.

%%%
\begin{widetext}
As an illustration of the utility of the operator formulation, we give explicit results for some  $\Omega_\kappa^{(1,3)}$s. These results could be used to build models that follow the structure in QCD, or perhaps someday to compute these matrix elements on the lattice.   For $\Omega_{qg \to q}^{(1)}$ and  $\Omega_{qg \to q}^{(3)}$, and the case where $c_e=g_e(r')=1$, we have
\begin{align}  \label{eq:O13qg}
\Omega_{qg \to q}^{(1)} &= \Omega_q^{(1)} = \int_0^1\!\!\df r\: \frac{1}{N_c}  \big\langle 0 \big| \tr \big\{ 
   Y_{\bar J}^\dagger Y_J \hat \cE_\perp(r)    Y_J^\dagger Y_{\bar J}  \big\} \big| 0 \big\rangle 
  \,,
  \\
\Omega_{qg \to q}^{(3)} &= \int_0^1\!\!\df r\: \frac{(-1)}{N_c} \Big\langle 0\, \Big| \tr\Big\{  
  \Big[Y_{\bar J}^\dagger Y_J \hat \cE_\perp(r)    Y_J^\dagger Y_{\bar J}, 
  \frac{1}{{\bar n_J} \sdt \img \partial}\,  g {n_J} \sdt \cB_{\bar J} 
    + \frac{2}{({\bar n_J} \sdt \img \partial)^2}\,\img \partial_\perp \sdt g \cB_{{\bar J}\perp} 
    - \frac{1}{{\bar n_J} \sdt \img \partial} \Big[\frac{1}{{\bar n_J} \sdt \img \partial} g \cB_{{\bar J}\perp}^\mu, g \cB_{{\bar J}\perp\mu} \Big]\Big]
\nn \\ & \qquad
+\Big[\Big[Y_{\bar J}^\dagger Y_J \hat \cE_\perp(r)    Y_J^\dagger Y_{\bar J},
  \frac{1}{{\bar n_J} \sdt \img \partial}\, g  \cB_{{\bar J}\perp}^\mu\Big],\frac{1}{{\bar n_J} \sdt \img \partial}\, g  \cB_{{\bar J}\perp\mu}\Big]
+\frac{1}{C_F} \Big[\Big[Y_{\bar J}^\dagger Y_J \hat \cE_\perp(r)    Y_J^\dagger Y_{\bar J},
  \frac{1}{{\bar n_J} \sdt \img \partial}\, g \cB_{{\bar J}\perp}^{\mu A}\Big],\frac{1}{{\bar n_J} \sdt \img \partial}\, g \cB_{{\bar J}\perp\mu}^A\Big]
\nn \\ & \qquad 
- \frac{(1-r^2)}{2} Y_{\bar J}^\dagger Y_J \hat \cE_\perp(r)    Y_J^\dagger Y_{\bar J}
\Big\} \Big|\,0 \Big\rangle
 \nn
\,,\end{align}
%%%
where the Wilson lines $Y_J$, $Y_{\bar J}$, and ${\cal B}_{\bar J}^\nu = {\cal B}_{\bar J}^{\nu A} T^A= \frac{1}{g} [ Y_{\bar J}^\dagger \img D^\nu Y_{\bar J}]$ are all in the fundamental representation, and $\tr$ is a trace over $3$ and $\bar 3$ color indices. The path for $Y_J^\dagger Y_{\bar J}$ is  $[-\infty,0]$ along $\bar n_J$, then $[0,\infty]$ along $n_J$. The measurement is normalized such that $\hat \cE_\perp(r) = 2\pi \hat \cE_\perp(r,0,0)$, which is equal to $\hat \cE_\perp(r,y=0)$ of Ref.~\cite{Mateu:2012nk}.  In \eq{O13qg} the inverse derivatives $1/(\bar n_J\sdt\img\partial)$ only act on the fields they are next to, and the fields on the right (left) side of the measurement $\hat \cE_\perp(r)$ are (anti) time-ordered. For $\Omega_{q\bar q \to g}^{(1)}$, $\Omega_{gg \to g}^{(1)}$, and $\Omega_{q\bar q \to g}^{(3)}$, and the case where $c_e=g_e(r')=1$, we have
%%%
\begin{align}  \label{eq:O13qqb}
\Omega_{q\bar q \to g}^{(1)} &= \Omega_g^{(1)} 
  =  \int_0^1\!\!\df r\: \frac{1}{N_c^2\!-\! 1}  \big\langle 0 \big| {\rm Tr} \big\{ 
   \cY_{\bar J}^T \cY_J \hat \cE_{\perp}\!(r)    \cY_J^T \cY_{\bar J}  \big\} \big| 0 \big\rangle 
   = \Omega_{gg\to g}^{(1)}  \,,
  \\
\Omega_{q\bar q \to g}^{(3)} &=\!\int_0^1\!\!\!\df r\:  \frac{(-1)}{N_c^2\!-\! 1}  \Big\langle 0 \,\Big| {\rm Tr} \Big\{
\Big[ \cY_{\bar J}^T \cY_J  \hat \cE_\perp\!(r)  \cY_{J}^T \cY_{\bar J} , 
  \frac{1}{{\bar n_J} \sdt \img \partial}\, g n_J \sdt \tilde \cB_{\bar J} 
  \!+\!\frac{2}{({\bar n_J} \sdt \img \partial)^2}\,\img \partial_\perp\! \sdt g \tilde \cB_{{\bar J}\perp} 
  \!+\!\frac{1}{{\bar n_J} \sdt \img \partial} \Big[ \frac{1}{{\bar n_J} \sdt \img \partial} g \tilde \cB_{{\bar J}\perp}^\mu, g \tilde   \cB_{{\bar J}\perp\mu} \Big] \Big]
\nn \\ & \quad
+\Big[\Big[ \cY_{\bar J}^T \cY_J  \hat \cE_\perp\!(r)  \cY_{J}^T \cY_{\bar J}  ,\frac{1}{{\bar n_J} \sdt \img \partial}\, g \bar \cB_{{\bar J}\perp}^\mu\Big],\frac{1}{{\bar n_J} \sdt \img \partial}\, g \bar \cB_{{\bar J}\perp\mu}\Big]
+ \frac{1}{T_F N_c} \Big[\Big[ \cY_{\bar J}^T \cY_J  \hat \cE_\perp\!(r)  \cY_{J}^T \cY_{\bar J}  ,\frac{1}{{\bar n_J} \sdt \img \partial}\, g \vec \cB_{{\bar J}\perp}^\mu\Big],\frac{1}{{\bar n_J} \sdt \img \partial}\, g \vec \cB_{{\bar J}\perp\mu}\Big]
\nn \\ & \qquad 
-\frac{(1-r^2)}{2} \cY_{\bar J}^T \cY_J  \hat \cE_\perp\!(r)  \cY_{J}^T \cY_{\bar J} 
\Big\} \Big|\, 0 \Big\rangle
\nn
\,,\end{align}
where the Wilson lines $\cY_J$ and $\cY_{\bar J}$ are in the adjoint representation, the gluon fields $\tilde \cB_{\bar J}^{ab} = -i f^{Cab} \cB_{\bar J}^C$ and $\bar \cB_{\bar J}^{ab} = d^{Cab} \cB_{\bar J}^C$ are matrices, and ${\rm Tr}$ is a trace over adjoint color indices.  When ${\rm Tr}$ acts on the term with $\vec \cB_{\bar J}^A=\cB_{\bar J}^A$ it simply contracts these color vectors to the appropriate sides of the color matrix $\cY_{\bar J}^T \cY_J  \hat \cE_\perp\!(r)  \cY_{J}^T \cY_{\bar J}$.
\end{widetext}
%%%

%===============================================================================
\subsection*{Soft function contribution to ISR}
%===============================================================================

\begin{figure}[t!]
	\hfill%
	\includegraphics[width=0.94\columnwidth]{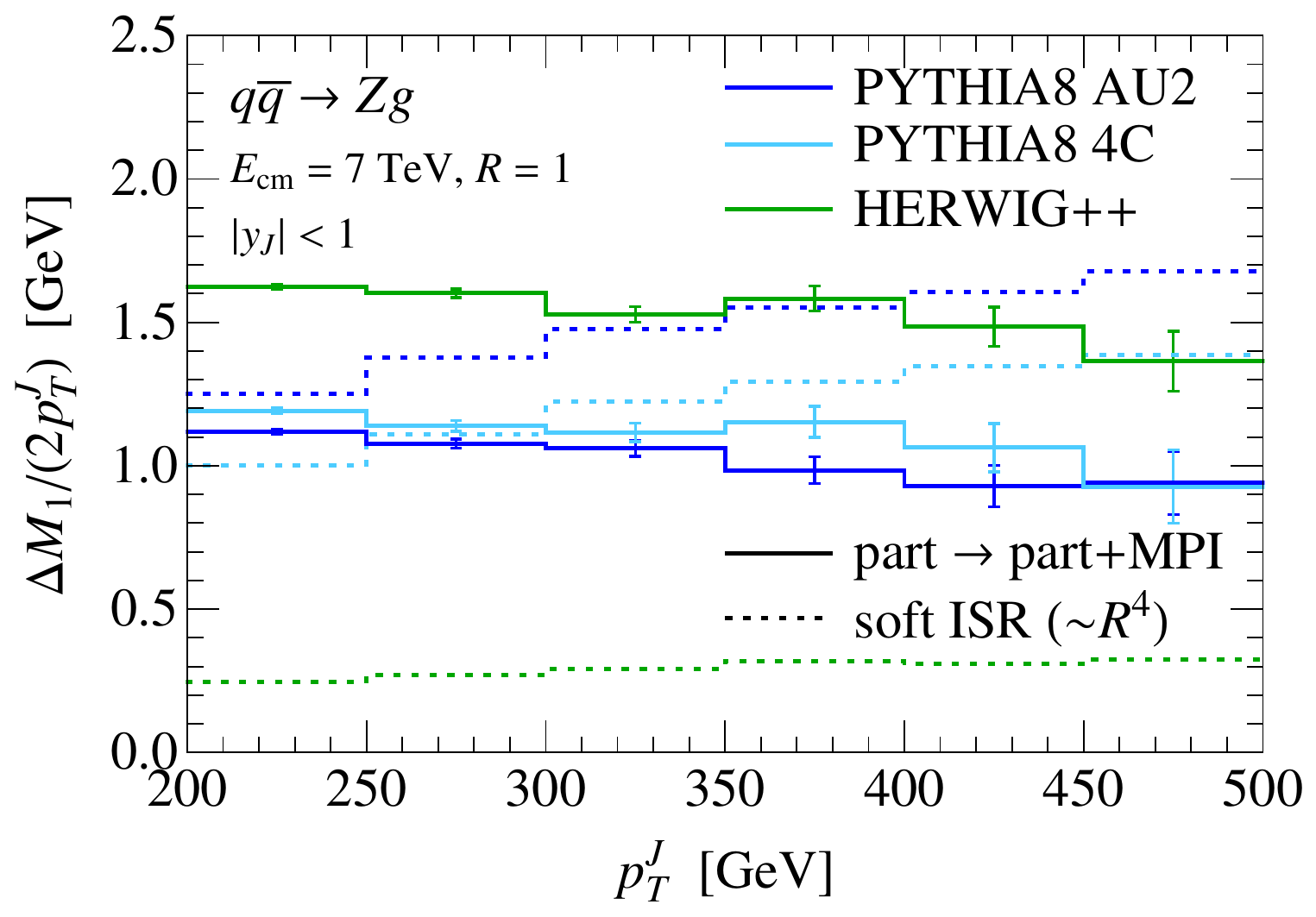}%
	\hspace*{\fill}
	\vspace{-2ex}
	\caption{Same as in \fig{DelM1MPI} but for $q\bar q\to Zg$. 
	}
	\label{fig:DelM1MPI2}
	\vspace{-1ex}
\end{figure}

\begin{figure*}[t!]
	\hfill%
	\includegraphics[width=0.9\columnwidth]{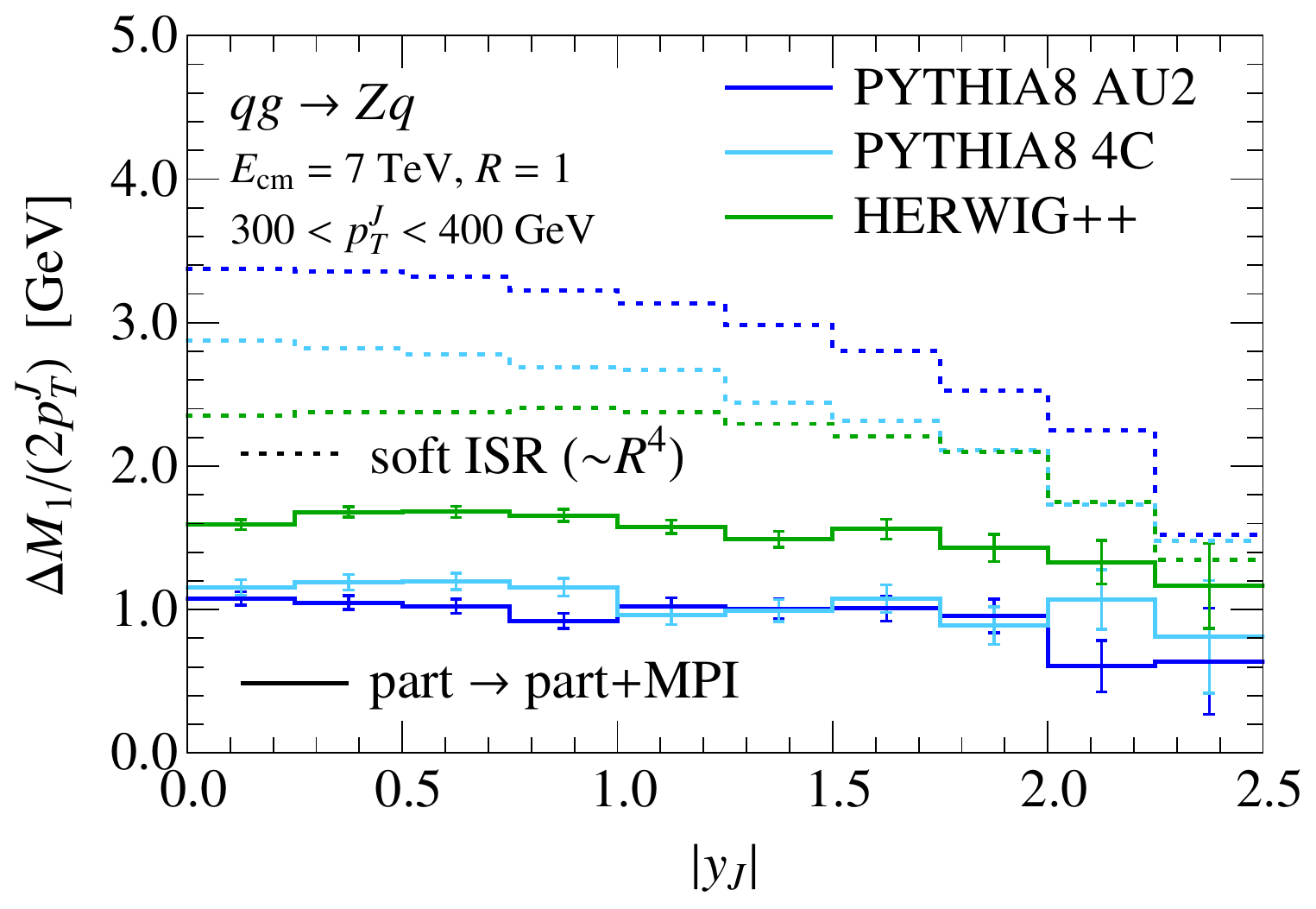}%
	\hfill\hfill%
	\includegraphics[width=0.9\columnwidth]{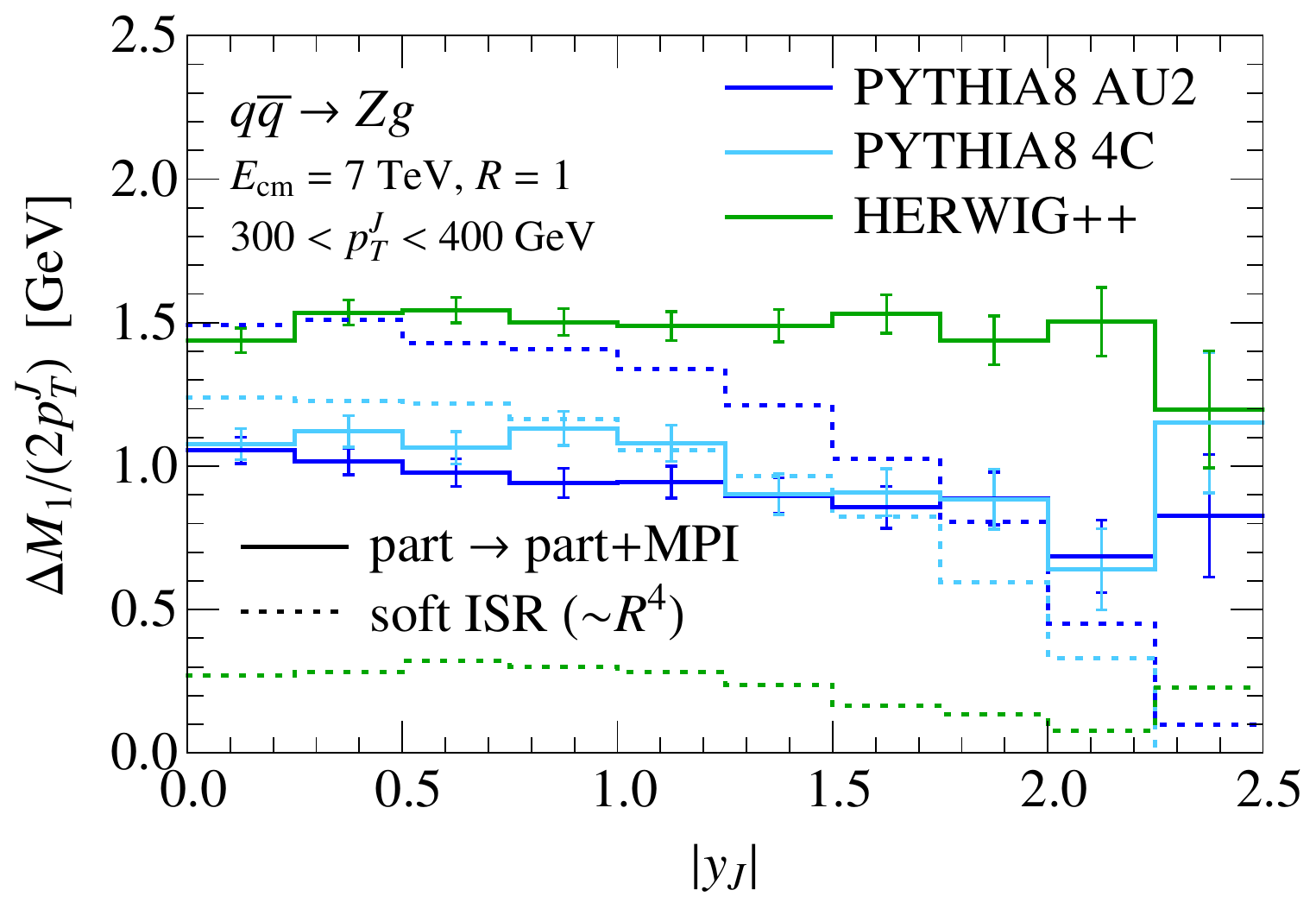}%
	\hspace*{\fill}
	\vspace{-2ex}
	\caption{Same as \fig{DelM1MPI} but for the $y_J$ dependence.}
	\label{fig:DelM1MPI_yJ}
	\vspace{4ex}
	\vspace{-1ex}
\end{figure*}

\begin{figure*}[t!]
	\hfill%
	\includegraphics[width=0.9\columnwidth]{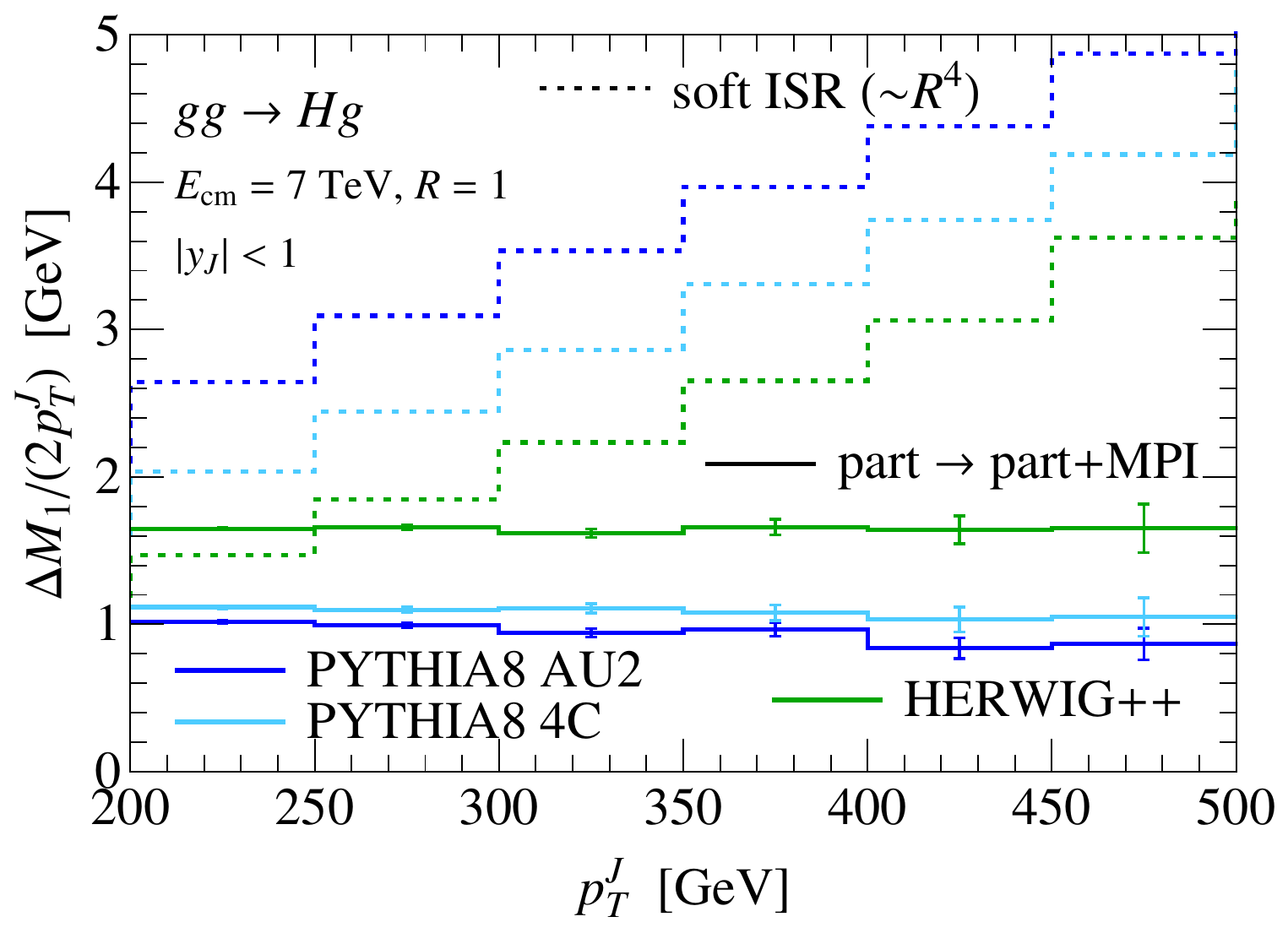}%
	\hfill\hfill%
	\includegraphics[width=0.9\columnwidth]{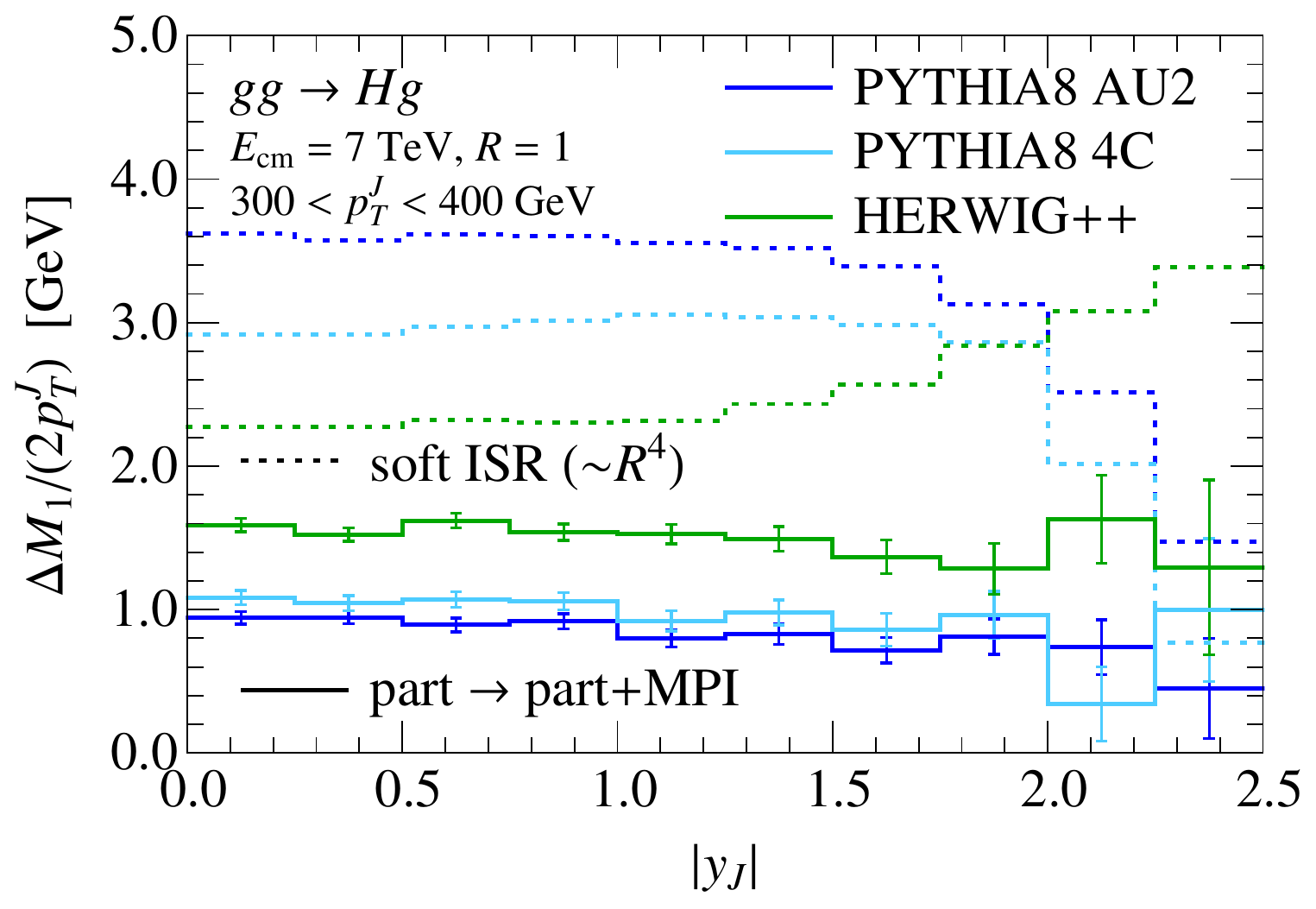}%
	\hspace*{\fill}
	\vspace{-2ex}
	\caption{Same as \fig{DelM1MPI} and \fig{DelM1MPI_yJ} but for the $gg\to Hg$ process.}
	\label{fig:DelM1MPI_Hg}
	\vspace{-1ex}
\end{figure*}

At $\ord{\al_s}$ the soft function contains the following term~\cite{Jouttenus:2011wh}
%%%
\begin{align} \label{eq:soft_interf}
  S^{\rm pert}_\kappa(k_S) \supset \bigl[I_0(\alpha,\beta) +I_0(\beta,\alpha)\bigr] \frac{\alpha_s C_\kappa}{\pi}\, \frac{1}{\mu} \Bigl(\frac{\mu}{k_S}\Bigr)_+
.\end{align}
%%%
The color factor for this interference of soft ISR from the two beams is given by the color charge of the two incoming partons, $C_\kappa = - \mathbf{T}_a \cdot \mathbf{T}_b$. For the processes we consider, this is simply a number given in \eq{ISRcolor}, but in general this is a matrix in color space.
The $I_0$ in \eq{soft_interf} is given by the following integral
%%%
\begin{align} \label{eq:I0}
I_0(\alpha, \beta)
&=  \frac{1}{\pi}\int_{-\pi}^{\pi}\!\df\phi \int\!\df y\,
\theta\bigl(e^{y_J -y} - \sqrt{\beta/\alpha}\bigr)
\\ & \quad \times
\theta\bigl(1/\alpha - 1 - e^{2(y_J -y)} + 2 e^{y_J -y} \cos\phi\bigr)
\,.\nn\end{align}
%%%
with parameters 
\begin{align}
\alpha = (1- \tanh y_J)/(2\rho)
\,, \nn \\
\beta = (1+ \tanh y_J)/(2\rho) 
\,.\end{align}
%%%
Here, $\rho(R,y_J)$ controls the jet size, which is chosen such that the jet area in $(y,\phi)$ space equals $\pi R^2$~\cite{Jouttenus:2011wh}. The total integral in \eq{I0} is an area in $(y,\phi)$ space, where the second theta function restricts the integral to the jet and the first theta function reduces to $\theta(y<0)$ and $\theta(y>0)$ for  $I_0(\alpha,\beta)$ and $I_0(\beta,\alpha)$, respectively. Therefore, including the overal $1/\pi$ factor,
%%%
\begin{align}
I_0(\alpha,\beta) + I_0(\beta,\alpha) = R^2
\,,\end{align}
%%%
which yields the $R^2$ dependence shown in \eq{soft_interf_short}.

The $p_T$ dependence of the MPI and soft ISR contributions to the jet mass moment is discussed in \fig{DelM1MPI}. In \fig{DelM1MPI_yJ} we show in addition the $y_J$ dependence in the same way. The $y_J$ dependence of the MPI is essentially flat, except for perhaps a small reduction at large rapidities. Since soft ISR emissions are constant in rapidity, one would expect the soft ISR contribution to the moment to be independent of the jet rapidity at central rapidities. This agrees well with what is observed in \Herwig for $\lvert y_J\rvert \lesssim 1.5$, while for larger $y_J$ the soft ISR contribution reduces. As already observed before, \Pythiaeight has a larger soft ISR and smaller MPI contribution than \Herwig. In addition, the $y_J$ dependence of the soft ISR differs noticably between \Pythiaeight and \Herwig. Hence, measurements of the $y_J$ rapidity dependence can also provide constraints on the modelling of soft ISR in the Monte Carlo programs.

For completeness we have included the analogs of \fig{DelM1MPI}, for the $q\bar q \to Zg$ channel in \fig{DelM1MPI2}, for the rapidity dependence in  \fig{DelM1MPI_yJ}, and for the $gg \to Hg$ channel in \fig{DelM1MPI_Hg}. Note that the size of the $R^4$ contribution from soft ISR for the gluon channel is very similar to $qg \to Zq$ (at central rapidities). This might be surprising since this is a purely gluonic process, but it is in agreement with the prediction from the color factors in \eq{ISRcolor}.

\newpage
\clearpage

\end{document}